\documentclass[%
superscriptaddress,showpacs,amsfonts,
amsmath,amssymb,
aps,
prc,
twocolumn,
]{revtex4-2}

\usepackage{graphicx}
\usepackage{dcolumn}
\usepackage{bm}
\usepackage{hyperref}

\usepackage[mathlines]{lineno}

\begin{document}

\preprint{APS/123-QED}

\title{Polarization Observables $T$ and $F$ in the $\gamma p \to \pi^0 p$ Reaction at CLAS}

\newcommand*{\GLASGOW}{University of Glasgow, Glasgow G12 8QQ, United Kingdom}
\newcommand*{\GLASGOWindex}{1}
\affiliation{\GLASGOW}
\newcommand*{\SCAROLINA}{University of South Carolina, Columbia, South Carolina 29208}
\newcommand*{\SCAROLINAindex}{2}
\affiliation{\SCAROLINA}
\newcommand*{\GWUI}{The George Washington University, Washington, DC 20052}
\newcommand*{\GWUIindex}{3}
\affiliation{\GWUI}
\newcommand*{\FZJ}{Institute for Advanced Simulation (IAS-4), Forschungszentrum J{\"u}lich, 52425 J{\"u}lich, Germany}
\newcommand*{\FZJindex}{4}
\affiliation{\FZJ}
\newcommand*{\ANL}{Argonne National Laboratory, Argonne, Illinois 60439}
\newcommand*{\ANLindex}{5}
\affiliation{\ANL}
\newcommand*{\ASU}{Arizona State University, Tempe, Arizona 85287-1504}
\newcommand*{\ASUindex}{6}
\affiliation{\ASU}
\newcommand*{\CSUDH}{California State University, Dominguez Hills, Carson, CA 90747}
\newcommand*{\CSUDHindex}{7}
\affiliation{\CSUDH}
\newcommand*{\CANISIUS}{Canisius University, Buffalo, NY}
\newcommand*{\CANISIUSindex}{8}
\affiliation{\CANISIUS}
\newcommand*{\CMU}{Carnegie Mellon University, Pittsburgh, Pennsylvania 15213}
\newcommand*{\CMUindex}{9}
\affiliation{\CMU}
\newcommand*{\SACLAY}{IRFU, CEA, Universit'{e} Paris-Saclay, F-91191 Gif-sur-Yvette, France}
\newcommand*{\SACLAYindex}{10}
\affiliation{\SACLAY}
\newcommand*{\CNU}{Christopher Newport University, Newport News, Virginia 23606}
\newcommand*{\CNUindex}{11}
\affiliation{\CNU}
\newcommand*{\UCONN}{University of Connecticut, Storrs, Connecticut 06269}
\newcommand*{\UCONNindex}{12}
\affiliation{\UCONN}
\newcommand*{\DUQUESNE}{Duquesne University, 600 Forbes Avenue, Pittsburgh, PA 15282 }
\newcommand*{\DUQUESNEindex}{13}
\affiliation{\DUQUESNE}
\newcommand*{\FU}{Fairfield University, Fairfield CT 06824}
\newcommand*{\FUindex}{14}
\affiliation{\FU}
\newcommand*{\FERRARAU}{Universita' di Ferrara , 44121 Ferrara, Italy}
\newcommand*{\FERRARAUindex}{15}
\affiliation{\FERRARAU}
\newcommand*{\FIU}{Florida International University, Miami, Florida 33199}
\newcommand*{\FIUindex}{16}
\affiliation{\FIU}
\newcommand*{\FSU}{Florida State University, Tallahassee, Florida 32306}
\newcommand*{\FSUindex}{17}
\affiliation{\FSU}
\newcommand*{\GSIFFN}{GSI Helmholtzzentrum fur Schwerionenforschung GmbH, D-64291 Darmstadt, Germany}
\newcommand*{\GSIFFNindex}{18}
\affiliation{\GSIFFN}
\newcommand*{\INFNFE}{INFN, Sezione di Ferrara, 44100 Ferrara, Italy}
\newcommand*{\INFNFEindex}{19}
\affiliation{\INFNFE}
\newcommand*{\INFNGE}{INFN, Sezione di Genova, 16146 Genova, Italy}
\newcommand*{\INFNGEindex}{20}
\affiliation{\INFNGE}
\newcommand*{\INFNRO}{INFN, Sezione di Roma Tor Vergata, 00133 Rome, Italy}
\newcommand*{\INFNROindex}{21}
\affiliation{\INFNRO}
\newcommand*{\INFNTUR}{INFN, Sezione di Torino, 10125 Torino, Italy}
\newcommand*{\INFNTURindex}{22}
\affiliation{\INFNTUR}
\newcommand*{\INFNCAT}{INFN, Sezione di Catania, 95123 Catania, Italy}
\newcommand*{\INFNCATindex}{23}
\affiliation{\INFNCAT}
\newcommand*{\INFNPAV}{INFN, Sezione di Pavia, 27100 Pavia, Italy}
\newcommand*{\INFNPAVindex}{24}
\affiliation{\INFNPAV}
\newcommand*{\ORSAY}{Universit'{e} Paris-Saclay, CNRS/IN2P3, IJCLab, 91405 Orsay, France}
\newcommand*{\ORSAYindex}{25}
\affiliation{\ORSAY}
\newcommand*{\KSU}{King Saud University, Riyadh, Kingdom of Saudi Arabia}
\newcommand*{\KSUindex}{26}
\affiliation{\KSU}
\newcommand*{\KNU}{Kyungpook National University, Daegu 41566, Republic of Korea}
\newcommand*{\KNUindex}{27}
\affiliation{\KNU}
\newcommand*{\LAMAR}{Lamar University, 4400 MLK Blvd, PO Box 10046, Beaumont, Texas 77710}
\newcommand*{\LAMARindex}{28}
\affiliation{\LAMAR}
\newcommand*{\MIT}{Massachusetts Institute of Technology, Cambridge, Massachusetts  02139-4307}
\newcommand*{\MITindex}{29}
\affiliation{\MIT}
\newcommand*{\MISS}{Mississippi State University, Mississippi State, MS 39762-5167}
\newcommand*{\MISSindex}{30}
\affiliation{\MISS}
\newcommand*{\UNH}{University of New Hampshire, Durham, New Hampshire 03824-3568}
\newcommand*{\UNHindex}{31}
\affiliation{\UNH}
\newcommand*{\NSU}{Norfolk State University, Norfolk, Virginia 23504}
\newcommand*{\NSUindex}{32}
\affiliation{\NSU}
\newcommand*{\OHIOU}{Ohio University, Athens, Ohio  45701}
\newcommand*{\OHIOUindex}{33}
\affiliation{\OHIOU}
\newcommand*{\ODU}{Old Dominion University, Norfolk, Virginia 23529}
\newcommand*{\ODUindex}{34}
\affiliation{\ODU}
\newcommand*{\JLUGiessen}{II Physikalisches Institut der Universitaet Giessen, 35392 Giessen, Germany}
\newcommand*{\JLUGiessenindex}{35}
\affiliation{\JLUGiessen}
\newcommand*{\RPI}{Rensselaer Polytechnic Institute, Troy, New York 12180-3590}
\newcommand*{\RPIindex}{36}
\affiliation{\RPI}
\newcommand*{\URICH}{University of Richmond, Richmond, Virginia 23173}
\newcommand*{\URICHindex}{37}
\affiliation{\URICH}
\newcommand*{\ROMAII}{Universita' di Roma Tor Vergata, 00133 Rome Italy}
\newcommand*{\ROMAIIindex}{38}
\affiliation{\ROMAII}
\newcommand*{\SDU}{Shandong University, Qingdao, Shandong 266237, China}
\newcommand*{\SDUindex}{39}
\affiliation{\SDU}
\newcommand*{\MSU}{Skobeltsyn Institute of Nuclear Physics, Lomonosov Moscow State University, 119234 Moscow, Russia}
\newcommand*{\MSUindex}{40}
\affiliation{\MSU}
\newcommand*{\TEMPLE}{Temple University,  Philadelphia, PA 19122 }
\newcommand*{\TEMPLEindex}{41}
\affiliation{\TEMPLE}
\newcommand*{\JLAB}{Thomas Jefferson National Accelerator Facility, Newport News, Virginia 23606}
\newcommand*{\JLABindex}{42}
\affiliation{\JLAB}
\newcommand*{\ULS}{Universidad de La Serena, Avda. Juan Cisternas 1200, La Serena, Chile}
\newcommand*{\ULSindex}{43}
\affiliation{\ULS}
\newcommand*{\UTFSM}{Universidad T\'{e}cnica Federico Santa Mar\'{i}a, Casilla 110-V Valpara\'{i}so, Chile}
\newcommand*{\UTFSMindex}{44}
\affiliation{\UTFSM}
\newcommand*{\BRESCIA}{Universit`{a} degli Studi di Brescia, 25123 Brescia, Italy}
\newcommand*{\BRESCIAindex}{45}
\affiliation{\BRESCIA}
\newcommand*{\UCR}{University of California Riverside, 900 University Avenue, Riverside, CA 92521, USA}
\newcommand*{\UCRindex}{46}
\affiliation{\UCR}
\newcommand*{\UTK}{University of Tennessee, Knoxville, Tennessee 37996, USA}
\newcommand*{\UTKindex}{47}
\affiliation{\UTK}
\newcommand*{\YORK}{University of York, York YO10 5DD, United Kingdom}
\newcommand*{\YORKindex}{48}
\affiliation{\YORK}
\newcommand*{\VIRGINIA}{University of Virginia, Charlottesville, Virginia 22901}
\newcommand*{\VIRGINIAindex}{49}
\affiliation{\VIRGINIA}
\newcommand*{\YEREVAN}{Yerevan Physics Institute, 375036 Yerevan, Armenia}
\newcommand*{\YEREVANindex}{50}
\affiliation{\YEREVAN}

\newcommand*{\NOWJLAB}{Thomas Jefferson National Accelerator Facility, Newport News, Virginia 23606}
\newcommand*{\NOWISU}{Idaho State University, Pocatello, Idaho 83209}
\newcommand*{\NOWCUA}{Catholic University of America, Washington, D.C. 20064}

\author {H.~Jiang} 
\affiliation{\GLASGOW}
\author {S.~Strauch} 
\affiliation{\SCAROLINA}
\author {I.I.~Strakovsky} 
\affiliation{\GWUI}
\author {D. R{\"o}nchen} 
\affiliation{\FZJ}
\author {R.~Workman} 
\affiliation{\GWUI}
\author {P.~Achenbach} 
\affiliation{\JLAB}
\author {J. S. Alvarado} 
\affiliation{\ORSAY}
\author {H.~Atac} 
\affiliation{\TEMPLE}
\author {L.~Baashen} 
\affiliation{\KSU}
\author {N.A.~Baltzell} 
\affiliation{\JLAB}
% ...

\collaboration{The CLAS Collaboration}
\noaffiliation

\date{\today}

\begin{abstract}
Pion photoproduction in the $\overrightarrow\gamma\overrightarrow p \to \pi^0 p$ reaction has been measured in the FROST experiment at the Thomas Jefferson National Accelerator Facility. In this experiment, circularly polarized photons with energies up to 3.082 GeV impinged on a transversely polarized frozen-spin target. Final-state protons were detected in the CEBAF Large Acceptance Spectrometer. Results of the polarization observables $T$ and $F$ have been extracted for $W$ from 1445 MeV to 2525 MeV.  The data generally agree with predictions of present partial-wave analyses, but also show marked differences for higher $W$ ranges. By incorporating the present data into the databases, the SAID fits have been improved with relatively small $\chi^2$ and significant changes in the parameters of the $\Delta(1910)1/2^+$ and $N(2190)7/2^-$ have been found with the J{\"u}Bo model.
\end{abstract}

\maketitle


\section{Introduction}

Quantum Chromodynamics (QCD)  describes the strong interactions of quarks and gluons. Non-perturbative QCD works in the low energy regime and the study of baryons is ideal to understand the strong interaction in the low energy regime. The study of excited baryon resonances reveals the strong interaction in the quark confinement and provides complementary information on the structure of the nucleon~\cite{baryons}~\cite{mai2023towards}. Resonances may not have been observed due to a weak coupling of the missing states to the reaction channel of the experiment or may not exist because the degrees of freedom used in the quark model are incorrect. A more complete knowledge of resonances will improve our understanding of the underlying symmetries and quark-quark interactions as resonances reflect the dynamics and relevant degrees-of-freedom within hadrons.

To get better knowledge of quark wave functions inside a nucleon, meson photoproduction experiments are useful as they reveal the dynamics of the quarks and the nucleon excited states. The meson photoproduction experiments were summarized in Refs.~\cite{ireland2020photoproduction}~\cite{thiel2022light}. In addition to the mass and width values provided by elastic pion-nucleon scattering, the single-pion photoproduction process provides confirmations to the baryon excited states and determines the amplitudes. Information about baryon resonances can be obtained from the complex amplitudes of the single-pion photoproduction process, which can be extracted from the observables. The measurement of double-polarization observables with a polarized target is needed in addition to the unpolarized cross section and single-polarization observables, as they carry additional information about the complex amplitudes that does not exist in the unpolarized cross section and single-polarization observables. A summary of the definition of the polarization observables was available in Ref.~\cite{Barker:1975bp}. Some single-pion photoproduction observables were measured previously with CLAS at JLab, such as the observable $E$~\cite{201553} and $\Sigma$~\cite{dugger2013beam}. There were also measurements worldwide such as Refs.~\cite{HA2}~\cite{jermann2023measurement} from CBELSA/TAPS. In this analysis, the polarization observables $T$ and $F$ in the $\gamma p \to \pi^0 p$ reaction were extracted by using data from a polarized target experiment with CLAS~\cite{Keith201227}.

\begin{figure}[!htb]
\begin{center}
\includegraphics[width=3.4in]{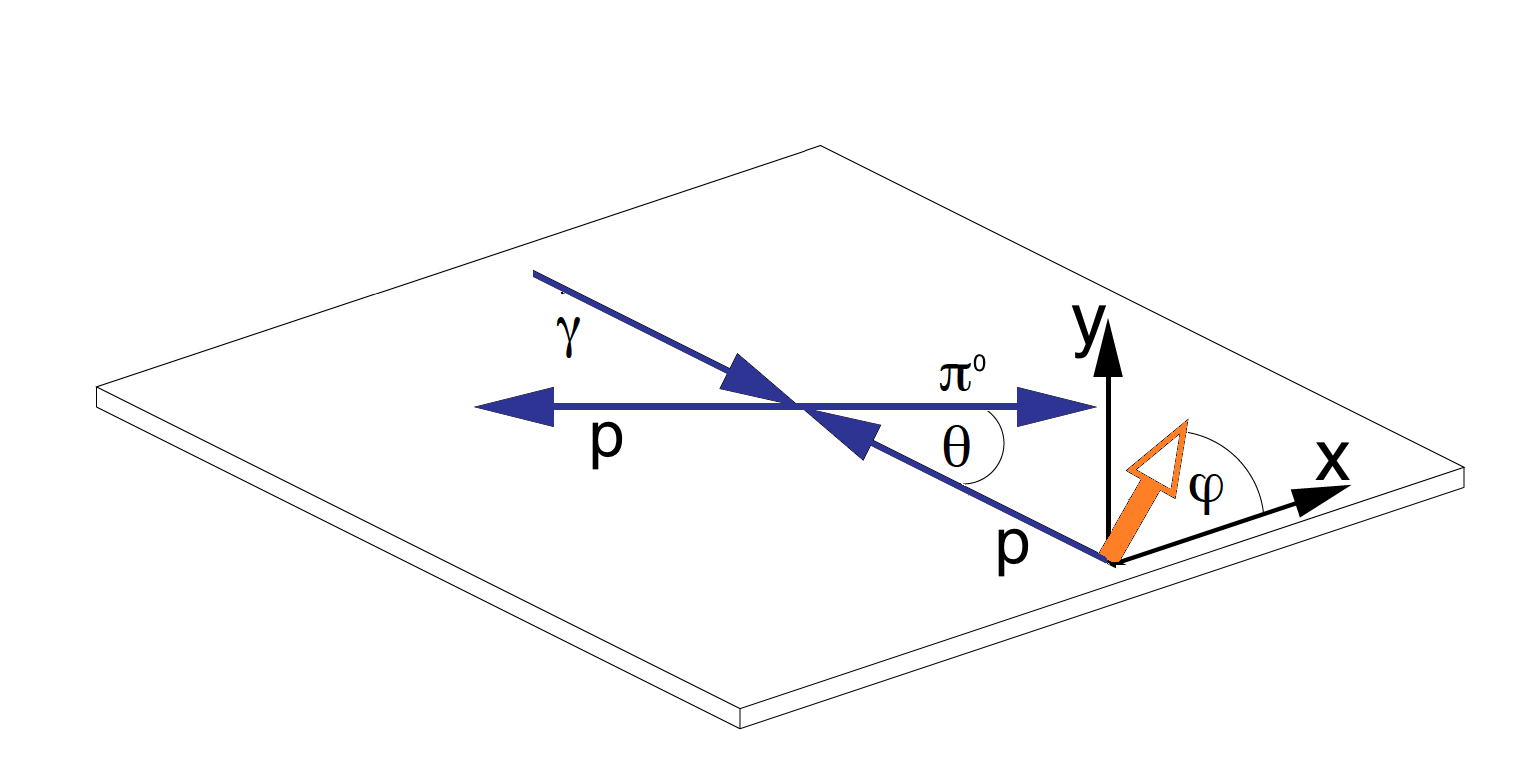}
\end{center}
\caption{Schematic of the $\gamma p \rightarrow \pi^0 p$ reaction in the center-of-momentum frame with circularly polarized photon beam and transversely polarized target. The target-polarization direction is indicated by the arrow with open arrow head.  \label{fig:angle}}
\end{figure}

The polarized cross section in single pseudoscalar-meson photoproduction with circularly polarized beam and transversely polarized target is given by~\cite{Barker:1975bp}
\begin{align}
\frac{d\sigma}{d\Omega} = 
\frac{d\sigma_0}{d\Omega}
 \left\{ 1 + P_T T \sin(\varphi) + P_T P_{\odot} F \cos(\varphi) \right\},
\label{eq:polcs}
\end{align}
where $\frac{d\sigma_0}{d\Omega}$ is the unpolarized cross section, $P_{\odot}$ is the degree of circular polarization of the photon beam, $P_T$ is the degree of target polarization, and $\varphi$ is the angle from the reaction plane to the target polarization direction.  All other polarization observables, including the proton-recoil polarizations, are fully integrated out even within a limited detector acceptance.  The schematic of the reaction is shown in Fig.~\ref{fig:angle}.
The coordinate system is defined as
\begin{align}
\hat x & = \hat y \times \hat z, \quad \\
\hat y & = \frac{\vec p^\text{\,cm}_\gamma \times \vec
         p^\text{\,cm}_{\pi^0}}{|\vec p^\text{\,cm}_\gamma \times \vec
         p^\text{\,cm}_{\pi^0}|}, \quad \text{and}\\
\hat z & = \frac{\vec p^\text{\,cm}}{|\vec p^\text{\,cm}|},
\end{align}
where $\vec p^\text{\,cm}_\gamma$ and $\vec p^\text{\,cm}_{\pi^0}$ are the momenta of the incident photon and final-state $\pi^0$ in the center-of-momentum frame.  The momentum of the proton is $\vec p^\text{\,cm}_p = -\vec p^\text{\,cm}_{\pi^0}$.

\section{Experimental Setup}

This experiment, E03-015 ``Pion Photoproduction from a Polarized Target'' \cite{e03105},  was conducted at the Thomas Jefferson National Accelerator Facility (Jefferson Lab)~\cite{jlab2}. Experimental data were taken between March 2010 and August 2010 using the CEBAF Large Acceptance Spectrometer (CLAS)~\cite{Mecking:2003zu} in Hall B at Jefferson Lab.

Longitudinally polarized electrons from the electron source were accelerated to 67 MeV and injected into the accelerator. The electrons were accelerated up to 6 GeV after five accelerating passes through pairs of antiparallel 600~MeV linear accelerators (LINACs) connected by recirculation arcs and delivered to the experimental halls~\cite{jlab2}. After receiving the accelerated electrons from the CEBAF accelerator, a polarized photon beam was produced by using the bremsstrahlung technique. The circularly polarized tagged photons with energies up to 3.082~GeV were produced by the bremsstrahlung radiator of the Hall-B Photon Tagger~\cite{Sober:2000we} from incident longitudinally polarized electrons.

The polarization of the photon beam depends on the ratio between the photon energy $E_\gamma$ and the electron beam energy $E_0$, $x=E_\gamma/E_0$. Specifically, the degree of the circular polarization is expressed as ~\cite{Olsen:1959zz}
\begin{align}
 P_{\odot} = P_e\frac{4x-x^2}{4-4x+3x^2}.
\label{eq:pgamma}
\end{align}
M{\o}ller measurements~\cite{moller} determined the electron-beam polarization. 
The average degree of electron-beam polarization and its statistical uncertainty are $P_e=0.873\pm 0.005$, while the systematic uncertainty of M{\o}ller measurements in Hall B was $3\%$~\cite{moller}. The electron-beam helicity was pseudo-randomly flipped at a rate of 240 Hz (30 Hz for the M{\o}ller measurements). 

The FROzen Spin Target (FROST) ~\cite{Keith201227} was used in the g9 experiment.  It was particularly designed for the experimental study of baryon resonances with the large acceptance CLAS detector.  Particle detection over a large polar angle range up to $135^{\circ}$ was permitted by the FROST target. The major components of FROST include the polarizing magnet, the dilution refrigerator, target material, and the holding coils.

The free protons polarized by dynamic nuclear polarization were kept at low temperature \cite{Keith201227}. At low temperatures, $\sim 30~\text{mK}$ for FROST, small holding fields are sufficient to maintain long spin-relaxation times.  The holding field direction can be aligned or anti-aligned with the proton polarization direction. The polarization time for the positive polarization was about 3400~h.  However, the relaxation time for the negative spin state was approximately half that of the positive \cite{Keith201227}. 
The collimated photon beam irradiated the FROST target. The butanol target material was positioned at the center of CLAS, $z=0$, and covered a range of 52.7~mm along the beamline. The nuclear spin of free protons in the target was transversally polarized, and the target polarization direction was changed periodically. In the target region, the photon beam also passed through the liquid helium coolant, the aluminum heat shield, as well as (2,2,6,6-Tetramethylpiperidin-1-yl)oxyl (TEMPO), water, and polychlorotrifluoroethylene (PCTFE) in the target and target insert. Additional 1.5-mm thick carbon and 3.5-mm thick polyethylene disks were mounted approximately 9~cm and 16~cm downstream of the butanol sample. These unpolarized targets helped clarify the bound-nucleon background and were used to study the instrumental asymmetries.

\begin{figure}[!htb]
\centering
\includegraphics[width=3.1in]{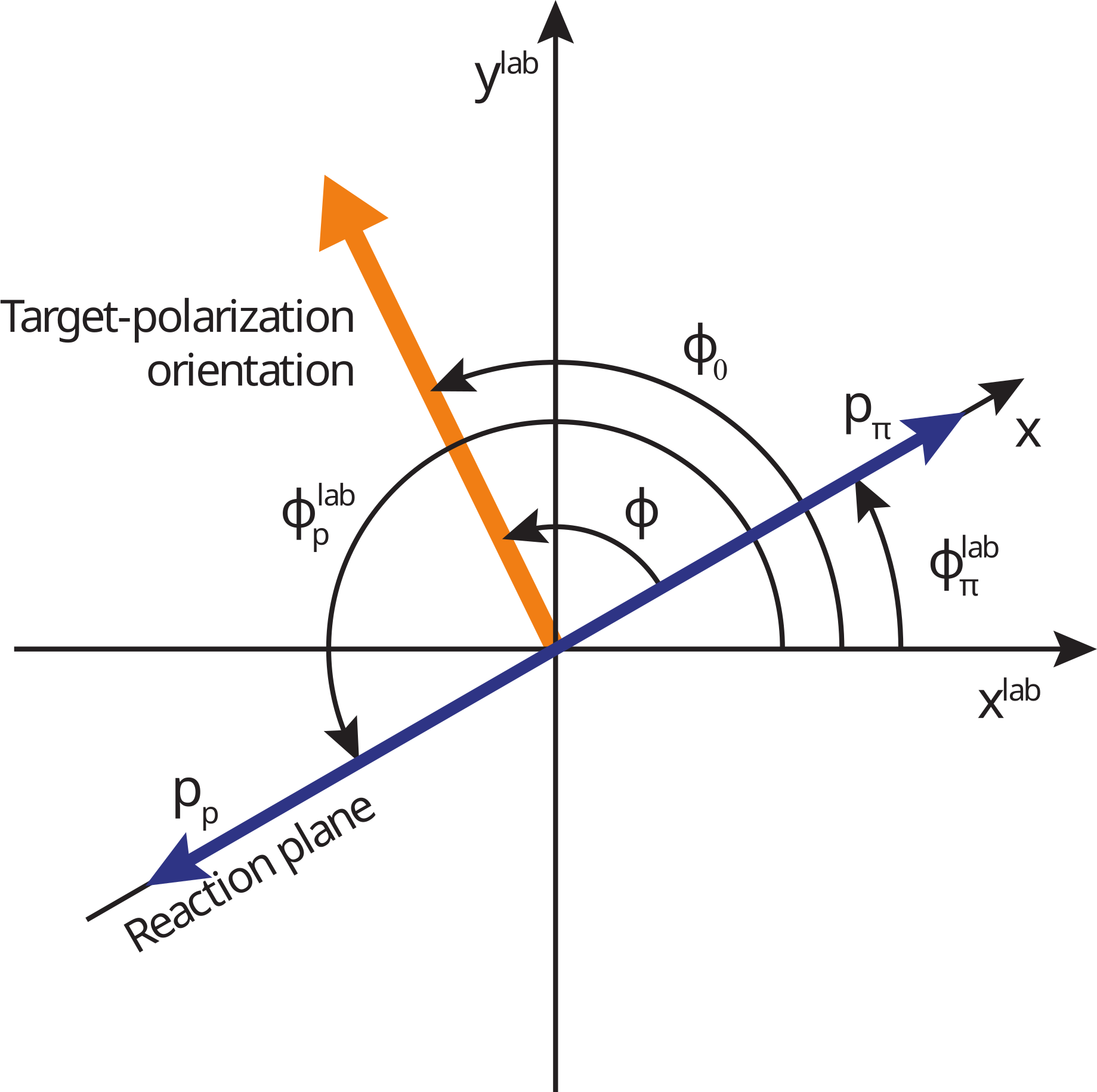}
\caption[Target polarization orientation in the lab and reaction frames]{Target polarization orientation in the lab and reaction frames.
\label{fig:tgt}}
\end{figure}

The target-polarization direction was oriented at an angle $\varphi_0$ with respect to the horizontal direction in the lab frame ($x^\text{lab}$) for the nominally positive polarization direction.  The angle $\varphi$ between the reaction plane and the target-polarization orientation, which enters in the polarized cross section Eq.~(\ref{eq:polcs}), was determined event by event from the measured azimuthal angle of the detected proton, $\varphi_p^\text{lab}$,
\begin{equation}
\varphi = -\varphi_\pi^\text{lab} + \varphi_0 = \pi -
\varphi_p^\text{lab} + \varphi_0.
\label{eq:phi}
\end{equation}
All relevant angles are illustrated in Fig.~\ref{fig:tgt}.  In this analysis a value of $\varphi_0 = (116.4 \pm 1.4)^\circ$ was used, which was determined in a moment-method analysis. A discussion of this is given in Sec. IV.B.

Nuclear Magnetic Resonance (NMR) measurements provided the degree of target polarization. The target-polarization values were determined in Ref.\cite{yuqing}.  The target polarization decreased with time and the target was routinely repolarized to the opposite direction after several runs.  In this analysis, it was assumed that the magnitude of the target polarization was constant within a given run.  A 4\% difference of measured FROST target polarizations with two separate NMR coils was reported in Ref.~\cite{Keith201227}. We take this difference as an estimate of the systematic uncertainty of the target-polarization values.

\begin{figure}[!htb]
\centering
\includegraphics[width=3.1in]{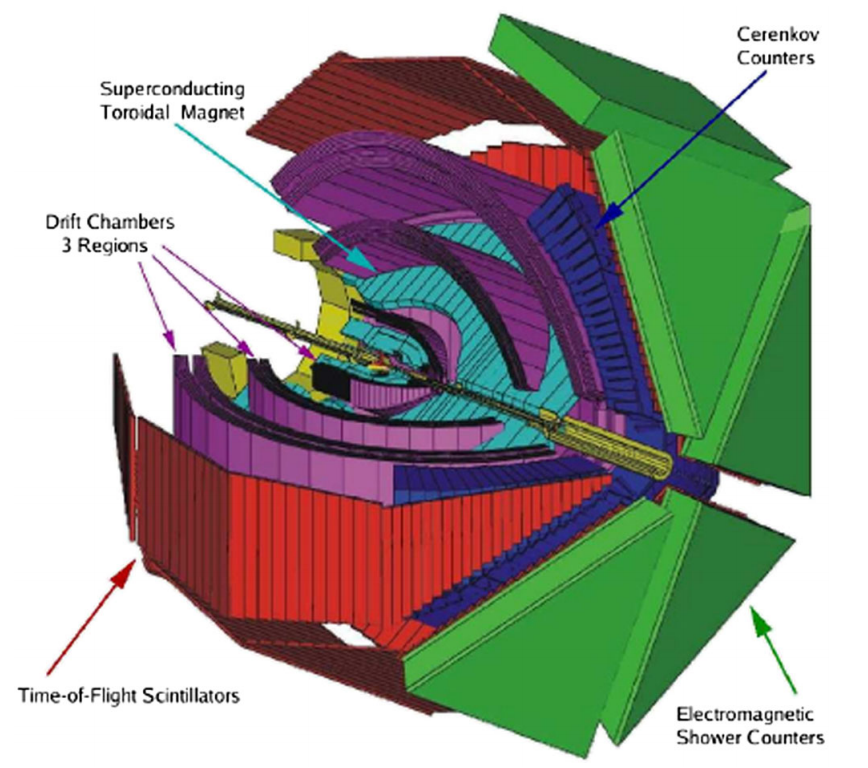}
\caption[Schematic view of the CLAS detector]{Schematic view of the CLAS detector, including drift chambers, Cherenkov counters, electromagnetic calorimeters, and time-of-flight counters. These detectors cover almost the entire $4\pi$ solid angle except the very forward and backward angles along the beamline. The detector measured about 9.14 m across.
\label{fig:clas}}
\end{figure}

The CLAS detector~\cite{Mecking:2003zu} was based on a multigap magnet with six super-conducting coils, symmetrically arranged to generate an approximately toroidal field distribution. The CLAS detector has been used for various experiments including the present pion photoproduction experiment.  One distinct aspect of the CLAS detector was the large-acceptance detection with a polar-angle range from $8^{\circ}$ to $142^{\circ}$.
It is particularly suited in the study of reactions with low luminosity (e.g., experiments using a tagged-bremsstrahlung photon beam) or reactions with multi-particle final states. 
The CLAS detector had several major components, including the torus magnet, the time-of-flight counters, the Cherenkov counters, and the drift chambers. The schematic view of the CLAS detector is shown in Fig.~\ref{fig:clas}. This design made it possible to detect final-states particles in a wide angular range.

The run information for all five data sets (run-groups) used in this analysis is summarized Table~\ref{tab:g9b}, which lists the run-group number, the run-number range, the number of events, the number of carbon-target events after applied analysis cuts, the frequency of the helicity flip, the target polarization, and the orientation of the target holding field~\cite{runinfo}. The incident electron-beam energy for all runs listed was 3.082 GeV.

\begin{table*}
\newcommand{\cc}[1]{\multicolumn{1}{c}{#1}}
\caption{\label{tab:g9b}%
 The g9b run information for runs with circularly polarized beam used in this analysis. Given are the run-group number, the range of run numbers, the number of identified $\gamma p \to pX$ events in the missing-mass range of a missing pion for the butanol target $N^B$ and for the carbon target $N^C$, the frequency of the helicity flip $f$, the target polarization and the sign of the target polarization, and the orientation of the target holding field~\cite{runinfo}.}

\begin{ruledtabular}
\begin{tabular}{ccdddcccccc}
Group & Run Range & \multicolumn{1}{c}{\textrm{$N^B~(10^6)$}} & \multicolumn{1}{c}{\textrm{$N^C~(10^6)$}} & f \text{ (Hz)} &  Target Pol. &  Field\\
\colrule
1 & 62207 -- 62289 & 0.943 & 0.726 & 240 & 0.83 -- 0.80 $(+)$ & $(+)$ \\
2 & 62298 -- 62372 & 1.641 & 1.239 & 240 & 0.86 -- 0.80 $(-)$ & $(+)$ \\
3 & 62374 -- 62464 & 1.675 & 1.245 & 240 \text{ or } 30 & 0.79 -- 0.75 $(+)$ & $(+)$ \\
4 & 62504 -- 62604 & 2.743 & 2.056 & 240 & 0.81 -- 0.76 $(-)$ & $(-)$ \\
5 & 62609 -- 62704 & 2.052 & 1.515 & 240 \text{ or } 30 & 0.85 -- 0.79 $(+)$ & $(-)$ \\
\end{tabular}
\end{ruledtabular}
\end{table*}

\section{Event Selection}

\subsection{Reaction Vertex}

\begin{figure}
\begin{center}
\includegraphics[width=3.4in]{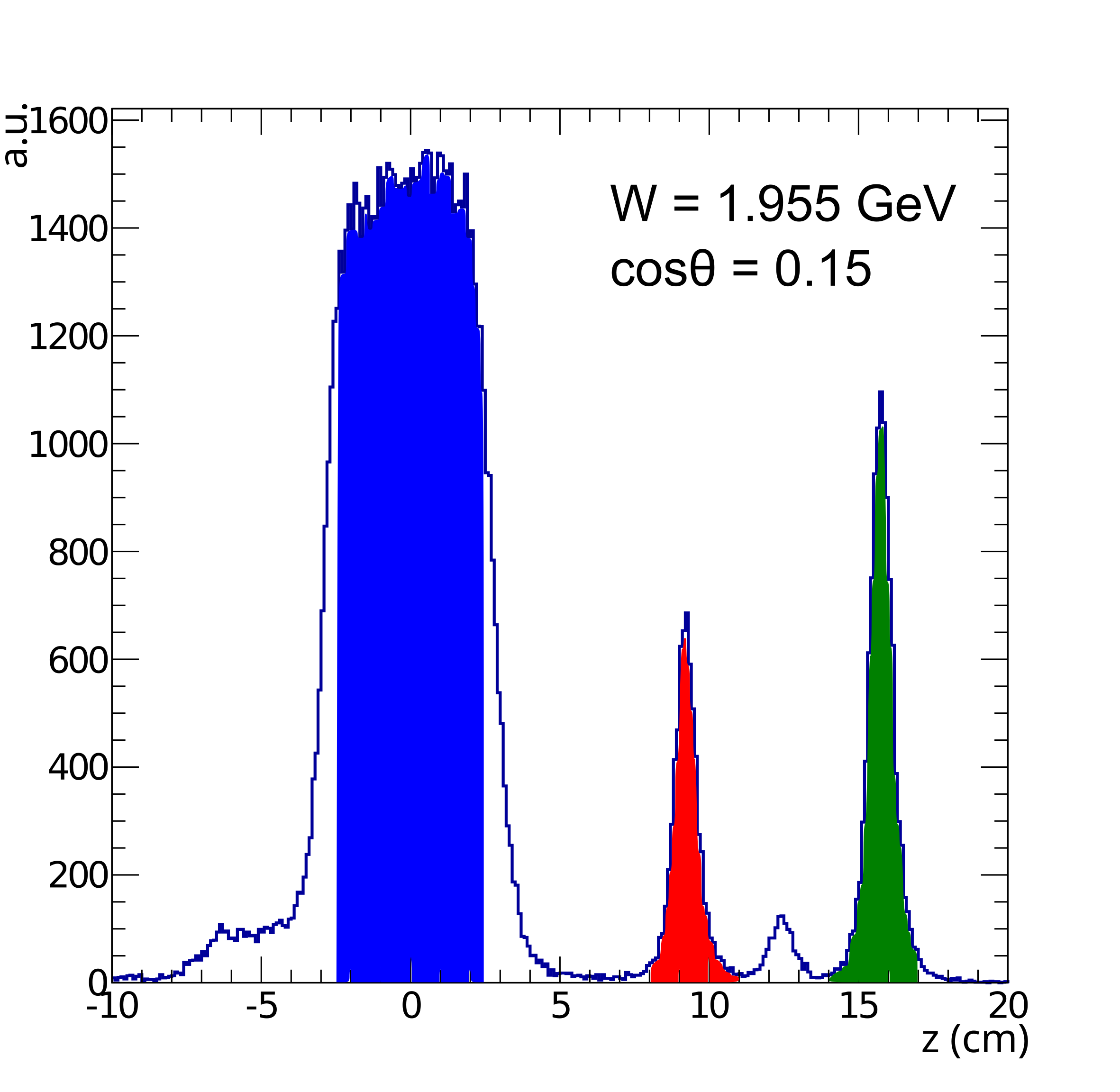}
\end{center}
\caption[The reconstructed $z$-vertex distribution of protons]{The reconstructed $z$-vertex distribution of protons. The main structures in the distribution are from the butanol ($z\approx 0$, marked with blue), carbon ($z \approx 9~\text{cm}$, marked with red), and polyethylene ($z \approx 16~\text{cm}$, marked with green) targets. According to the implementation of the target geometry in Ref.~\cite{momcor}, the structure at $z \approx 12.5~\text{cm}$ comes from the 1~K end cap.
  \label{fig:Z}}
\end{figure}

The reconstructed reaction vertices were utilized to categorize events from the butanol, carbon, and polyethylene targets in the beamline. A distribution of the $z$ coordinate of the reconstructed reaction vertex is shown in Fig.~\ref{fig:Z}. The selection range for butanol-target events, $z=-2.5$~cm to 2.5~cm, corresponds to the size of the 5.27-cm long target cup, centered at $z=0$.  The proton polarization of events reconstructed just outside that region drops quickly.
The figure illustrates the categorized targets in three colors based on the $z$ coordinate of the reconstructed reaction vertex.  The $z$-range of the butanol target was chosen to align with the highest average raw asymmetry from $z = -2.5$~cm to $z = 2.5$~cm.  The raw asymmetry of the observables $F$ starts to decrease beyond $z=2.5$~cm because the fraction of events from polarized protons starts to decrease as the reconstructed reaction vertex approaches the end of the actual target at $z \approx 2.5$~cm.  This cut helps to maintain a high average dilution factor.  A wider cut would have increased the overall statistics of the data but at the expense of a more diluted asymmetry signal.  The ranges of the applied $z$-vertex cuts for the three targets are listed in Table~\ref{tab:Z}.

\begin{table}[b]
\caption{\label{tab:Z}%
Selection criteria for the three targets based on the reconstructed $z$ coordinate of the reaction vertex.}
  \begin{ruledtabular}
\begin{tabular}{ccccccccc}
Target & Reaction Vertex ($z$) \\
\hline
Butanol & $-$2.5 cm to 2.5 cm \\
Carbon & 8 cm to 11 cm \\
Polyethylene & 14 cm to 17 cm \\
\end{tabular}
\end{ruledtabular}
\end{table}

A circular cut with a radius of 2 cm was applied on the transverse vertex coordinates to suppress poorly reconstructed proton tracks.

\subsection{Particle Identification and Coincidence Time}
\label{sec:pid}
The $\gamma p \to p \pi^0$ reaction was identified with the detection of one proton in coincidence with a bremsstrahlung photon and the reconstruction of a $\pi^0$ as the one unobserved particle in the final state.  First, events with one positively charged particle and zero negatively charged particles were considered.  The charge of a particle was determined by the curvature of its track in the magnetic field of CLAS.  Proton identification was then made by time-of-flight (TOF) from the reaction vertex to the TOF paddles. CLAS has a lower limit of proton momenta for a particle to be well reconstructed. Only proton candidates with momenta $p > 200$~MeV/$c$ were considered.  The time-of-flight difference of the positively charged particle in each event, $\Delta t_p$, was taken between the measured flight time, $t_{exp}$, and the flight time $t_{calc}$, which was calculated from the momentum $p$ and speed $\beta$ of the particle under the assumption that the particle was a proton with rest mass $m_p$:

\begin{equation}
\Delta t_{p} = t_{exp} - t_{calc} = \frac{\ell_{SC}}{c}\left(\frac{1}{\beta} - \sqrt{\frac{m_p^2c^2}{p^2}+1}\right),
\label{eq:pid}
\end{equation}
where $\ell_{SC}$ is the path length from the reaction vertex to the TOF paddles, and $\beta$ is determined from the time of flight and $\ell_{SC}$. $\Delta t_p$ is used to distinguish a proton from other positively charged particles.

\begin{figure}
\centering
\includegraphics[width=3.4in]{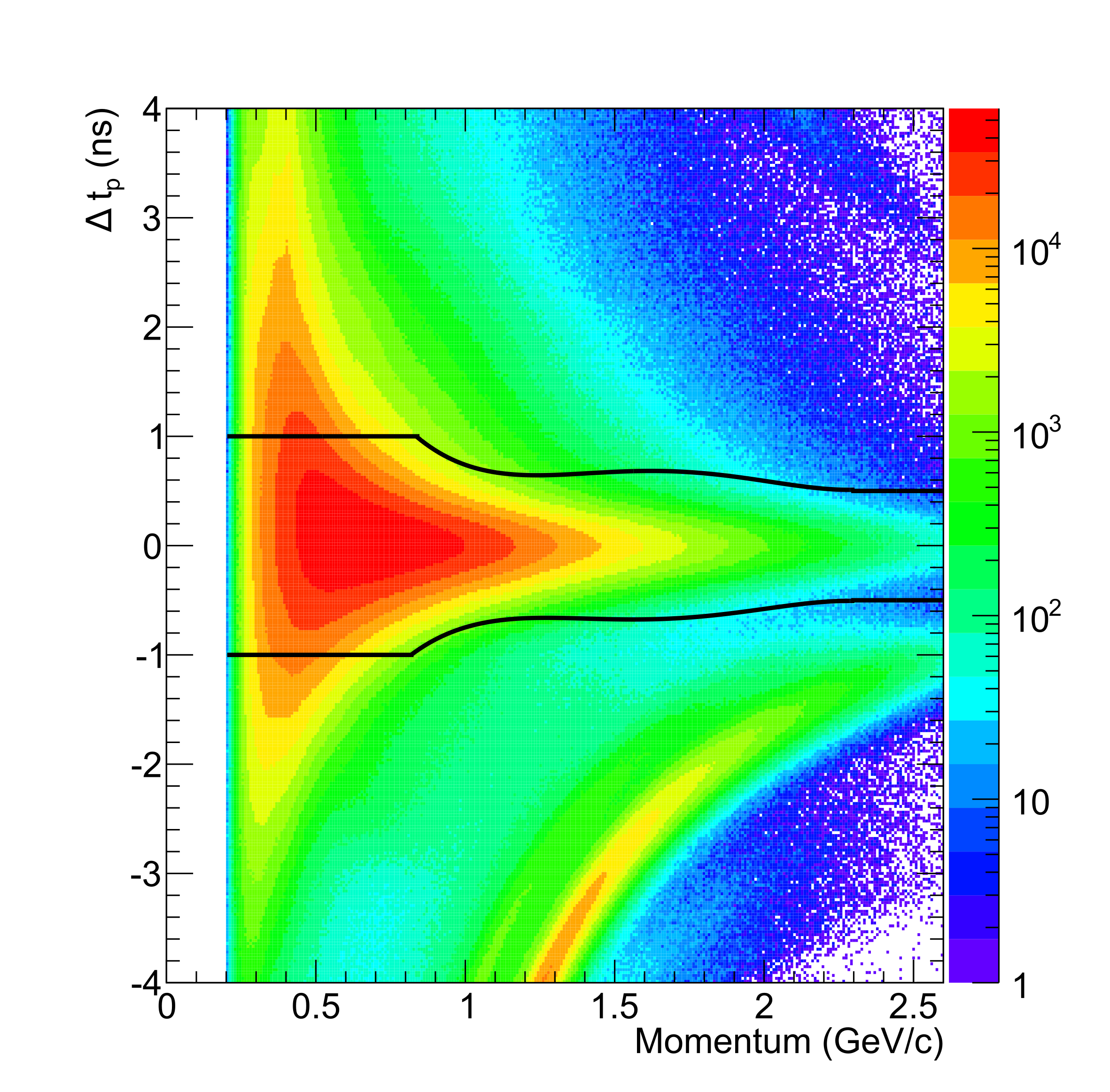}
\caption[Event selection cuts]{The $\Delta t_p$ distribution for positively charged particles in CLAS.  In this distribution, the black curves indicate the cut limits used to select the protons
   used in the analysis. \label{fig:events}}
\end{figure}

The distribution of $\Delta t_p$ for positively charged particles as a function of momentum is shown in Fig.~\ref{fig:events}.  The central maximum with $\Delta t_p \approx 0$ corresponds to protons, and the band at high momentum and negative values of $\Delta t_p$ corresponds to pions. 

The selection of the photon from all recorded photons that led to the reaction requires the CLAS-tagger coincidence time, $\Delta t_c$, which is defined as
\begin{equation}
\Delta t_{c} = t_{v,\gamma} - t_{v,p} =  \left(t_\gamma + \frac{z}{c}\right) - \left(t_{p,ST} - \frac{\ell_{ST}}{\beta c}\right),
\label{eq:coinc}
\end{equation}
where $t_\gamma$ is the photon time measured by the tagger and reported at the center of the CLAS detector, $z$ is the reaction vertex $z$ coordinate, $t_{p,ST}$ is the proton time of the Start Counter (ST) subsystem, and $\ell_{ST}$ is the path length of the proton from the reaction vertex to the hit position in the ST subsystem.

\begin{figure}
\centering
\includegraphics[width=3.4in]{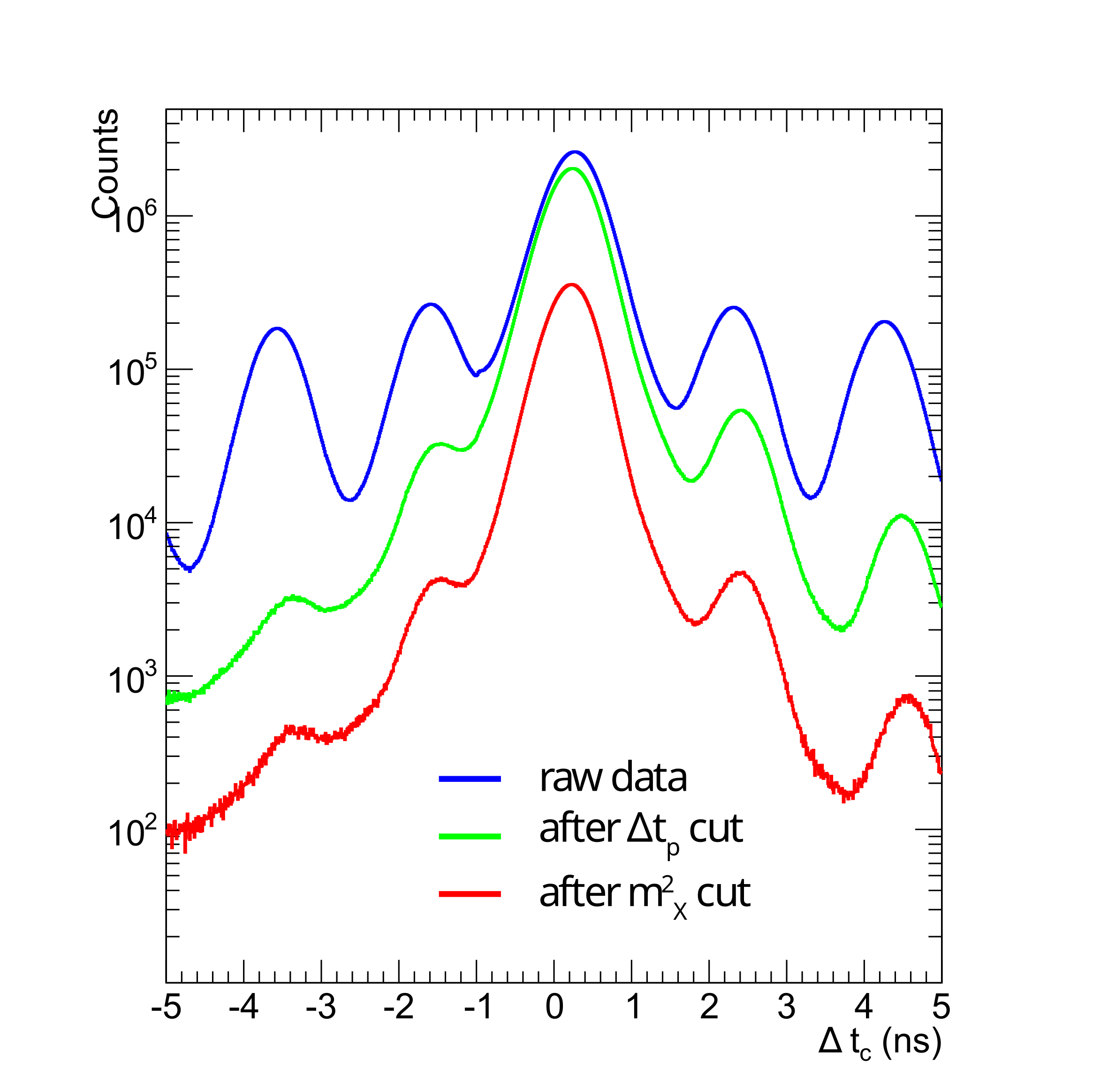}
\caption[CLAS-tagger coincidence time]{The CLAS-tagger coincidence time distribution for raw data (blue) and after the $\Delta t_p$ cut (green).  The red curve includes an additional cut on the missing mass close to the pion mass, $-0.04~\text{GeV$^2$/c$^4$} < m^2_X < 0.08~\text{GeV$^2$/c$^4$}$.
\label{fig:coin}}
\end{figure}

Event selections were made in either case with momentum-dependent cuts. To determine the cut ranges, the $\Delta t_c$ and $\Delta t_p$ distributions were sliced in 0.2-GeV/c wide momentum bins and fitted with Gaussian distributions.  The cut limits in the center of each momentum bin were chosen as $\pm 3\sigma$ off the distribution's maximum.  The individual cut limits were interpolated with a fourth-order polynomial to obtain the upper and lower limits of the momentum-dependent cut ranges.  The cut on $\Delta t_p$ was further restricted by $|\Delta t_p| < 1$~ns to reduce the amount of random background.  The loss of statistics below $p_p \approx 0.83$~GeV/c is minimal.  The amount of accidental background, however, is reduced by a factor of three with this additional cut.  For momenta $p_p > 2.3$~GeV/$c$, a cut $|\Delta t_p| < 0.5$~ns was imposed to minimize any possible proton misidentification with pions.  The final momentum-dependent event-selection regions are shown in Fig.~\ref{fig:events}.  Figure~\ref{fig:coin} shows the integrated $\Delta t_c$ distributions for various selection criteria: for the raw data, after applying the momentum-dependent $\Delta t_p$ cuts, and after selecting events with missing-mass-squared in the $\gamma p \to pX$ reaction close to $m^2_{\pi^0}$, $-0.04~\text{GeV$^2/c^4$} < m^2_X < 0.08~\text{GeV$^2/c^4$}$. This missing-mass-squared range corresponds on average to an approximately $3\sigma$ cut around the pion peak.  After the reaction-channel selection with a missing-mass cut, the background from random coincidences appears to be less than $2\%$ based on the yields in the true and random coincidence peaks.

From Fig.~\ref{fig:events} proton misidentification appears to be negligible. The central peak may have contamination up to 1\% from other peaks after cuts. The analysis accounts for potential particle misidentifications with $\sim 1\%$ in the uncertainty budget of the systematic uncertainties.

\subsection{Energy-Loss and Momentum Corrections}
\label{sec:corrections}
The standard {\tt eloss} package~\cite{momcor} was utilized to determine from the reconstructed momentum of the protons at the reaction vertex in the target.  The correction accounts for energy losses in the target material, target wall, carbon cylinder, and start counter. Additionally, momentum corrections were applied to the detected protons following the procedure of CLAS Note 2013-011, ``Momentum corrections for $\pi^+$ and protons in g9b data''~\cite{duggernote}.  These corrections attempt to correct errors in the momentum determination that may be caused by drift-chamber misalignments and uncertainties in the magnetic field map.

\subsection{Channel Identification}

\begin{figure*}
\centering
\includegraphics[width=3.1in]{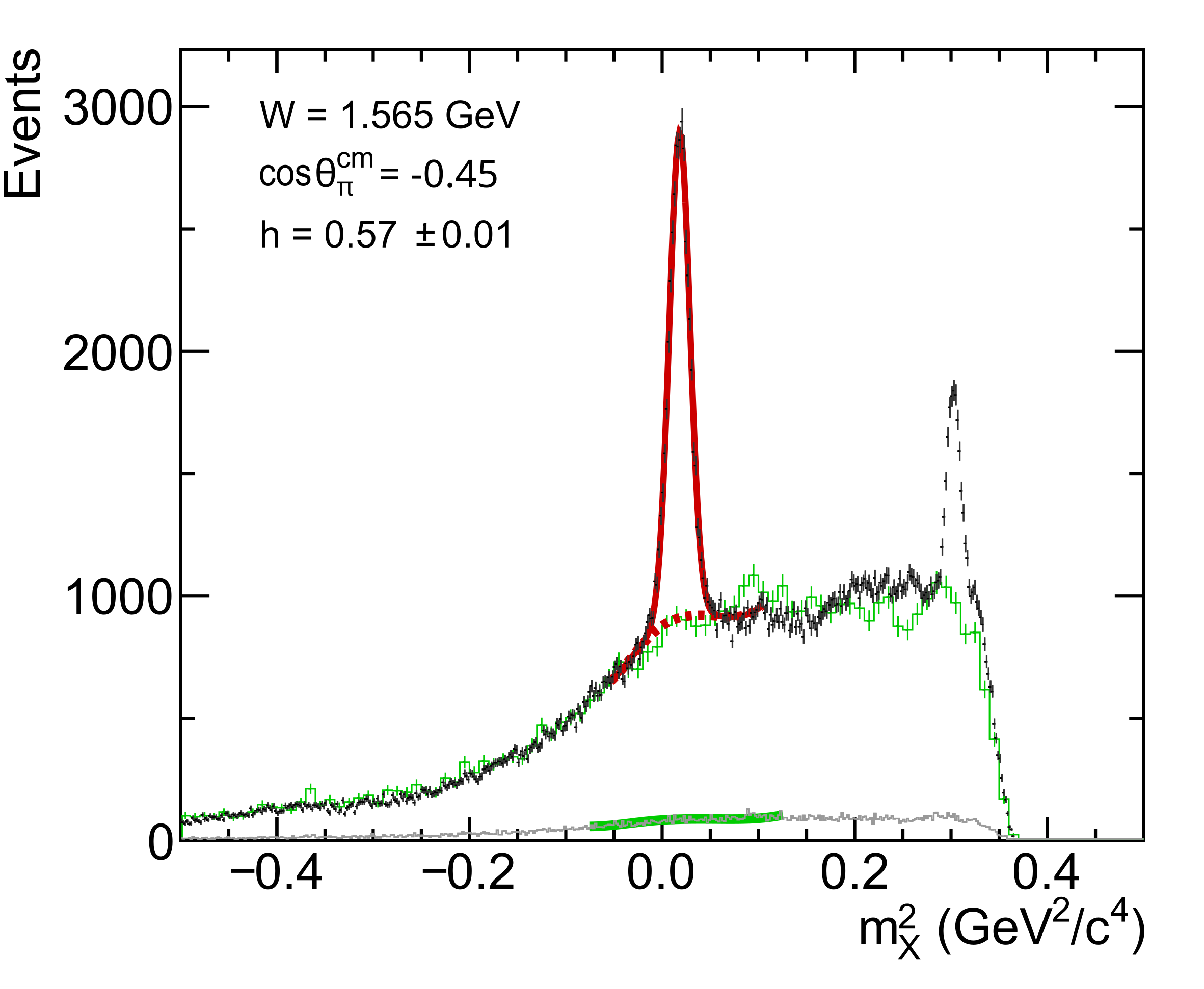}
\includegraphics[width=3.1in]{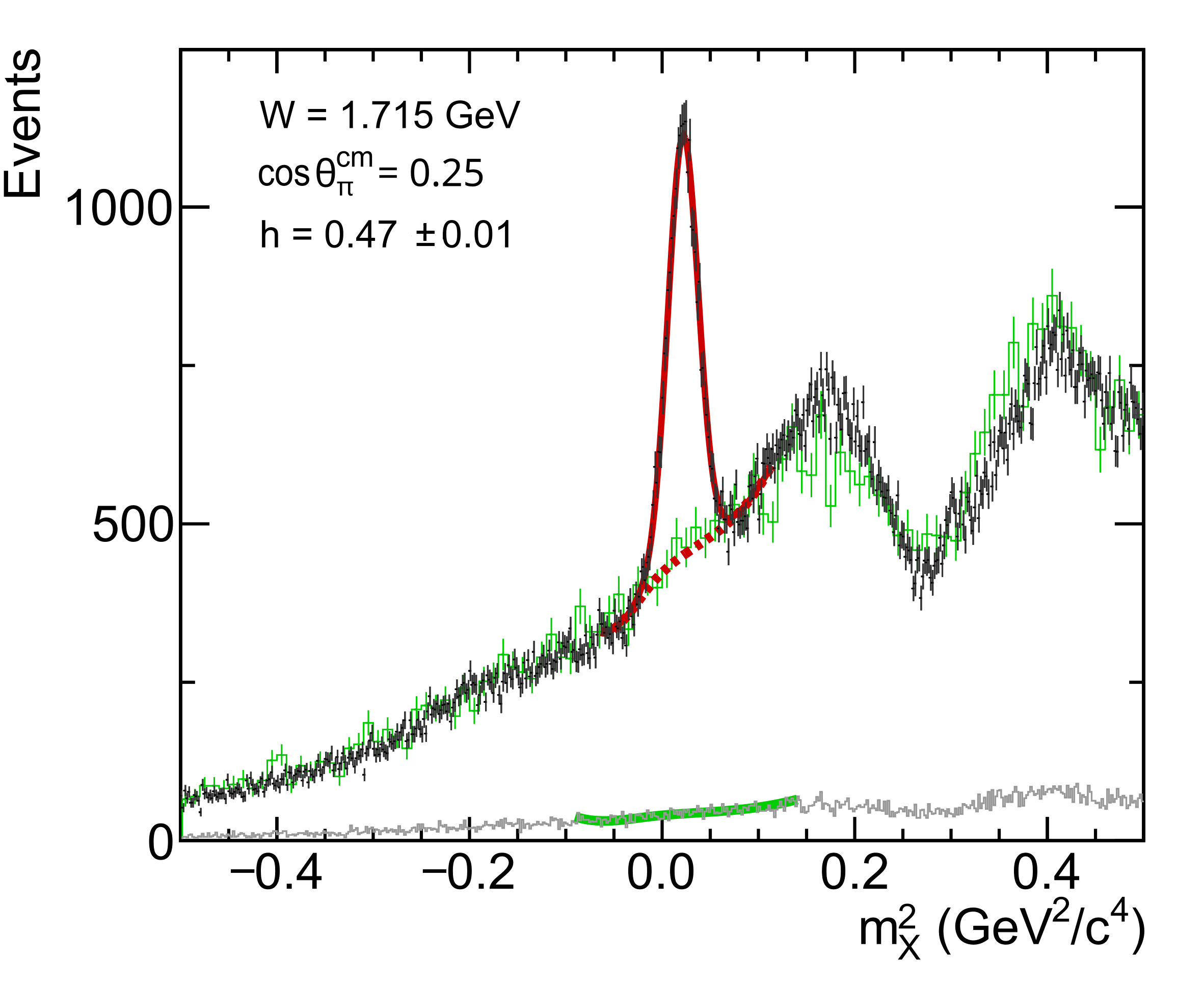}
\includegraphics[width=3.1in]{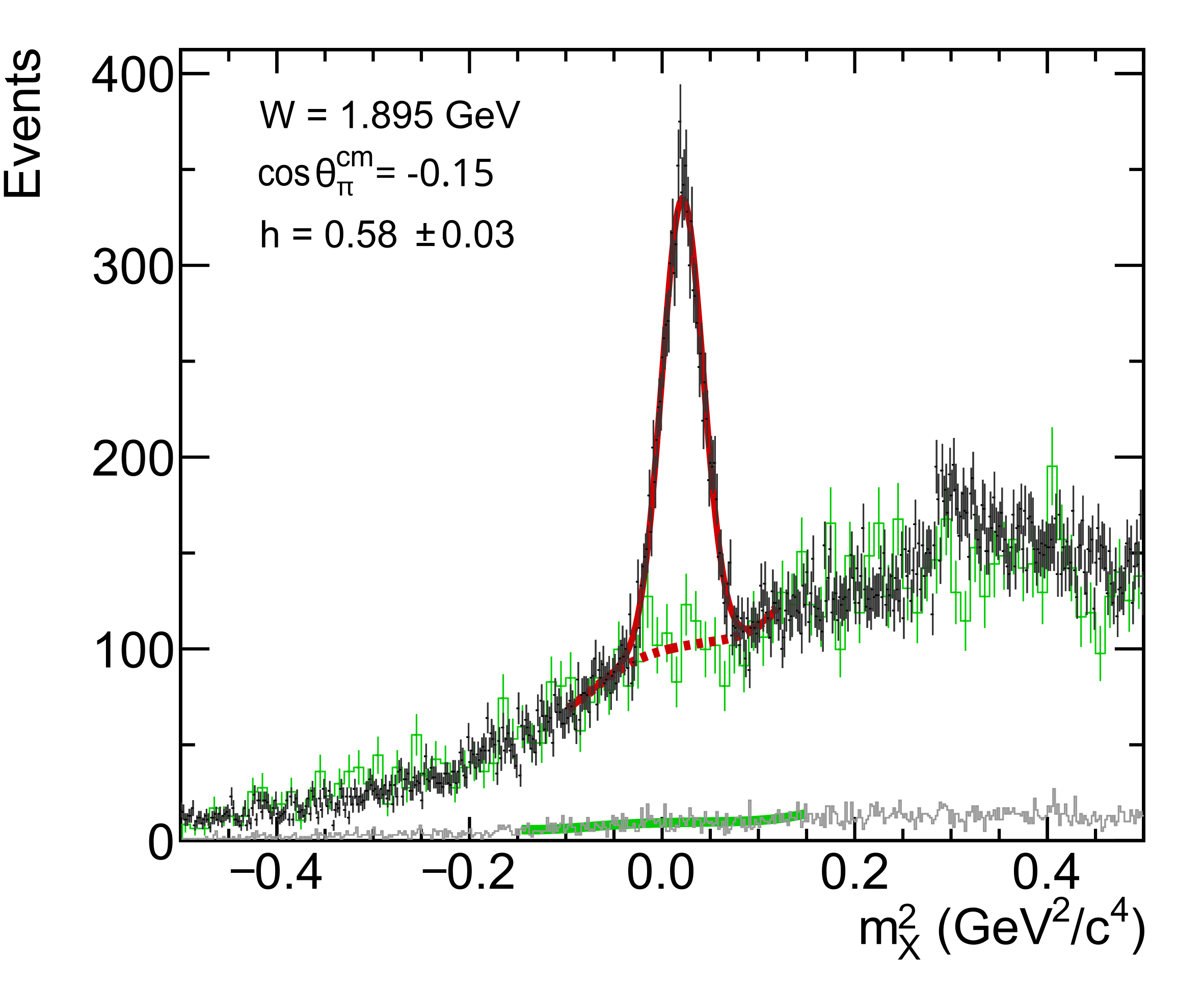}
\includegraphics[width=3.1in]{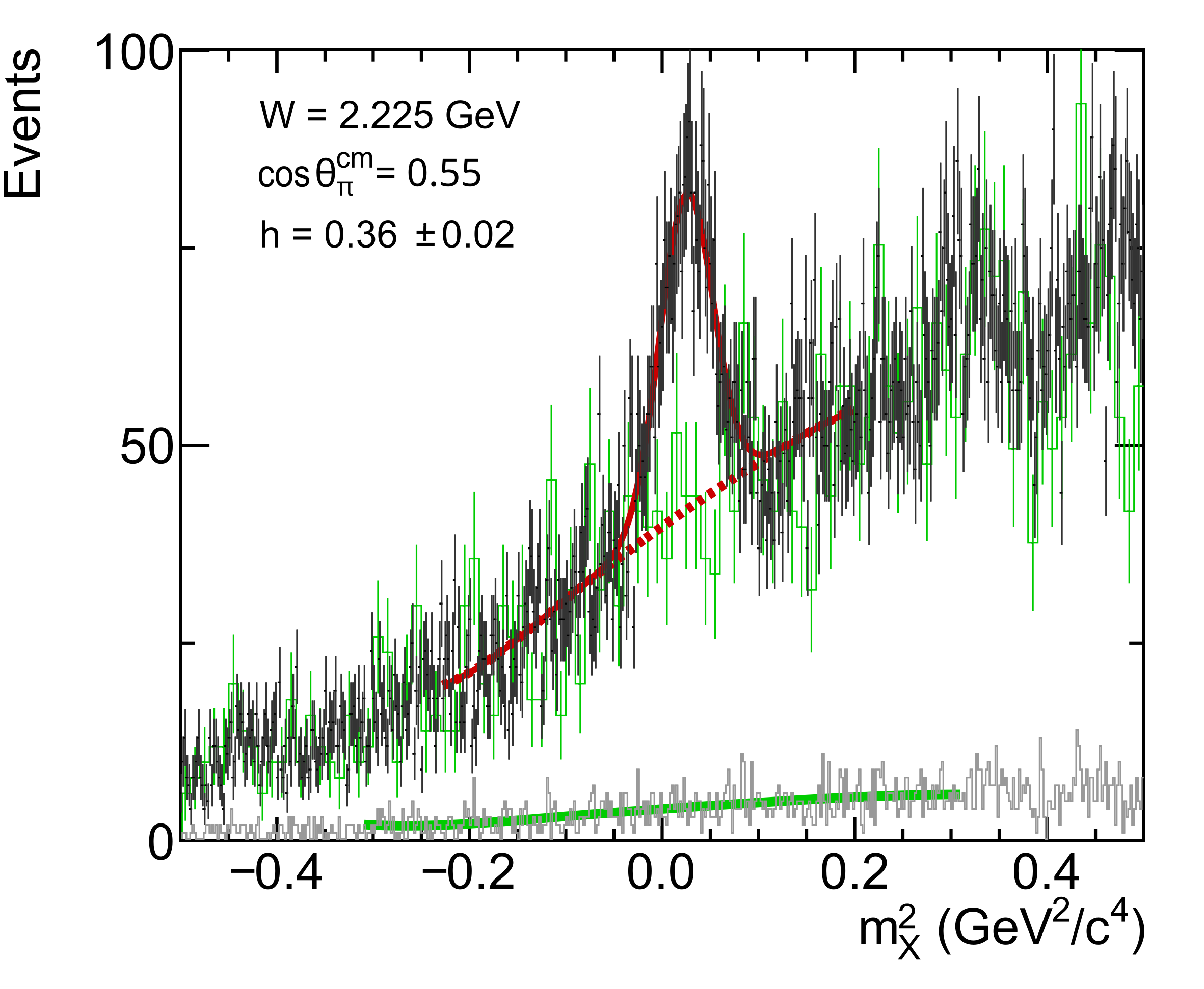}
\caption[Examples of missing-mass-squared distributions]{Examples of missing-mass-squared distributions from butanol- (black histograms) and carbon-target data (gray histograms).  The solid red and green curves are fits to the butanol and carbon data, respectively. The scaled carbon-data fits are shown as red dashed curves and the scaled carbon distributions as green histograms.  Also indicated are the dilution factors $h$. 
\label{fig:fit}}
\end{figure*}

The channel identification was based on individual kinematic bins. The data were sorted in 34 bins of $W$ from 1.49 GeV to 2.51 GeV with a bin size of 0.03~GeV and in 20 bins of $\cos\theta_{\pi}^{cm}$ from $-$1 to 1 with a bin size of 0.1. The missing-mass-squared distributions in the $\gamma p \to pX$ reaction from both the butanol and carbon targets were accumulated for each bin.  Figure~\ref{fig:fit} shows four examples of the missing-mass distributions for various energy and angular bins.

The black histograms show data from the butanol target and the gray histograms, with much smaller statistics, show data from the carbon target.  The central peaks at $m_X^2 = m_{\pi^0}^2 \approx 0.018~\text{GeV}^2/c^4$ above a broad background correspond to events from the $\gamma p \to \pi^0p$ reaction off free protons in the butanol target.  A function was fit to each distribution to determine the bound-nucleon background in the missing-mass-squared distributions from the butanol target. A well-separated $\eta$ peak can also be seen around 0.3 $\text{GeV}^2/c^4$.

For each bin, the carbon-target distribution was described with a cubic spline $p_3(x)$ with four nodes. The butanol-target distribution was described by the same spline function multiplied by a scale factor $\kappa$ for the bound-nucleon background and a Gaussian function for the free-proton signal events of the expected reaction channel.  The fit functions that described the signal, $S(x)$, the carbon-target, $C(x)$, and the butanol-target, $B(x)$, distributions are
\begin{align}
S(x) &= Y_0 \exp\left[ - \frac{(x - m^2_0)^2}{2\sigma^2}\right],  \label{eq:funcS}\\
C(x) &= p_3(x),\quad\text{and} \label{eq:funcC} \\
B(x) &= \kappa C(x) + S(x).
\end{align}

The signal parameters $Y_0$, $m^2_0$, and $\sigma$, the parameters of the spline function $x_i$, $y_i$, and the second derivative at the first and last nodes, $b_2$ and $e_2$, respectively, as well as the scale factor $\kappa$ were determined by simultaneous fits to the carbon and butanol distributions.  The quantity 
\begin{eqnarray}
  \chi_\text{MLE}^{2}=2\Big[\sum_i [B(m^2_i)-N^{B}_i\cdot \ln
  B(m^2_i)]\nonumber \\
+\sum_j[C(m^2_j)-N^{C}_j\cdot \ln C(m^2_j)]\Big]
\label{eq:chi}
\end{eqnarray}
was minimized in the fits.  It is the sum of the negative logarithm of the likelihood functions for Poisson-distributed butanol and carbon data.  The sums in each term are over both helicity states, all run-groups, and all azimuthal angles, effectively rendering the summed yields as unpolarized.  Minimizing $\chi_\text{MLE}$ introduces the least bias in the estimation of the parameters \cite{fowler2014maximum}.  The first sum in Eq.~(\ref{eq:chi}) runs over all bins in the butanol missing-mass-squared distribution within a specified fit range, listed in Table~\ref{tab:mm2}, and the second sum runs similarly over bins in the carbon-data histograms.  $N^B_k$ and $N^C_k$ are the number of events in the $k^\text{th}$ bins of the butanol and carbon histograms at a missing-mass-squared $m_k^2$, respectively.  The minimization of Eq.~(\ref{eq:chi}) was coded with the Minuit2 \cite{minuit2} package from CERN.

\begin{table*}
  \caption{Initial value for the peak width $\Delta m^2$ and missing-mass-squared ranges used for fits to the distributions from the butanol and carbon targets at the given energies.  The parameters for other values of $W$ are interpolated or taken as the range of the lowest (for $W < 1.5~\text{GeV}$) or highest (for $W > 2.4~\text{GeV}$) of the ranges given in the table.}
\label{tab:mm2}
\begin{ruledtabular}
\begin{tabular}{ccccccccc}
$W$ (GeV) & $\Delta m^2$  (GeV$^2/$c$^4$) & Butanol range (GeV$^2/$c$^4$) & Carbon range (GeV$^2/$c$^4$) \\
\hline
1.5 & 0.012 & $-$0.05 to 0.10 & $-$0.07 to 0.12 \\
1.8 & 0.016 & $-$0.07 to 0.12 & $-$0.10 to 0.15 \\
1.9 & 0.018 & $-$0.10 to 0.12 & $-$0.15 to 0.15 \\
2.0 & 0.020 & $-$0.15 to 0.15 & $-$0.15 to 0.20 \\
2.2 & 0.030 & $-$0.22 to 0.20 & $-$0.30 to 0.30 \\
2.4 & 0.035 & $-$0.30 to 0.20 & $-$0.40 to 0.40 \\
\end{tabular}
\end{ruledtabular}
\end{table*}

The fit ranges for the butanol-target data were chosen to ensure the coverage of the whole $\pi^0$ peak in the missing-mass-squared distribution and to keep away from structures of other reaction channels. The carbon-target distribution is more featureless, and a wider fit range was chosen to constrain the fit with these low-statistics data better. As the width of the $\pi^0$ peak increases with $W$, the fit ranges also increase. Examples of fit results are shown in Fig.~\ref{fig:fit}. The gross fit functions are shown as solid red curves and the scaled spline background functions are shown as dashed red curves. Only those $(W,\cos\theta_{\pi}^{cm})$ bins were selected for further analysis for which the $\pi^0$ peak and the background were clearly identified. Based on the histogram content and fit results, the following criteria were chosen for this selection:

\begin{itemize}
\item Number of events within $\pm \Delta m^2$ from $m^2_{\pi^0}$ in
  the butanol missing-mass distribution $N^B > 300$,
\item Scale factor: $\kappa + \sigma_\kappa > 8.0$ and $\kappa - \sigma_\kappa <
  12.5$,
\item Peak position:  $|m^2_0 - m^2_{\pi^0}| < 0.020$~ GeV$^2/c^4$,
\item Peak width: $\sigma > 0.006$~(GeV$^2/$c$^4$) and, for $W <
  2$~GeV, $\sigma < 0.030$~(GeV$^2/$c$^4$),
\item Dilution factor: $h + \sigma_h > 0.20$, and
\item Fit quality: $\chi_\text{MLE}^2 / \text{d.o.f} < 2.0$.
\end{itemize}

After obtaining the butanol missing-mass-squared distribution and the polynomial background for each kinematic bin, butanol-target events were selected that satisfy the $\gamma p \to \pi^0 p$ kinematics for further analysis.  The selection was based on the condition
\begin{align}
 |m^2_X-m^2_0| < n\sigma_H,
\label{eq:select}
\end{align}
where $m^2_0$ and $\sigma_H$ were the fit parameters in Eq.~(\ref{eq:funcS}), and $n=2$ was chosen for the final analysis.  For those events, normalized polarized yields and dilution factors were calculated.  The dilution factor is the ratio between the number of free proton events and the total number of events.  Experimentally, the dilution factor was determined from the number $N^B$ of selected butanol-target events, and the integral of the carbon-background function, Eq.~(\ref{eq:funcC}), over the missing-mass-squared range that satisfies the $\gamma p \to \pi^0 p$ kinematics in the specific kinematic bin, Eq.~(\ref{eq:select}),
\begin{equation}
h = 1 - \frac{\kappa}{N^B}\int C(m^2)dm^2.
\end{equation}

\begin{figure}[!htb]
  \centering
\includegraphics[width=3.4in]{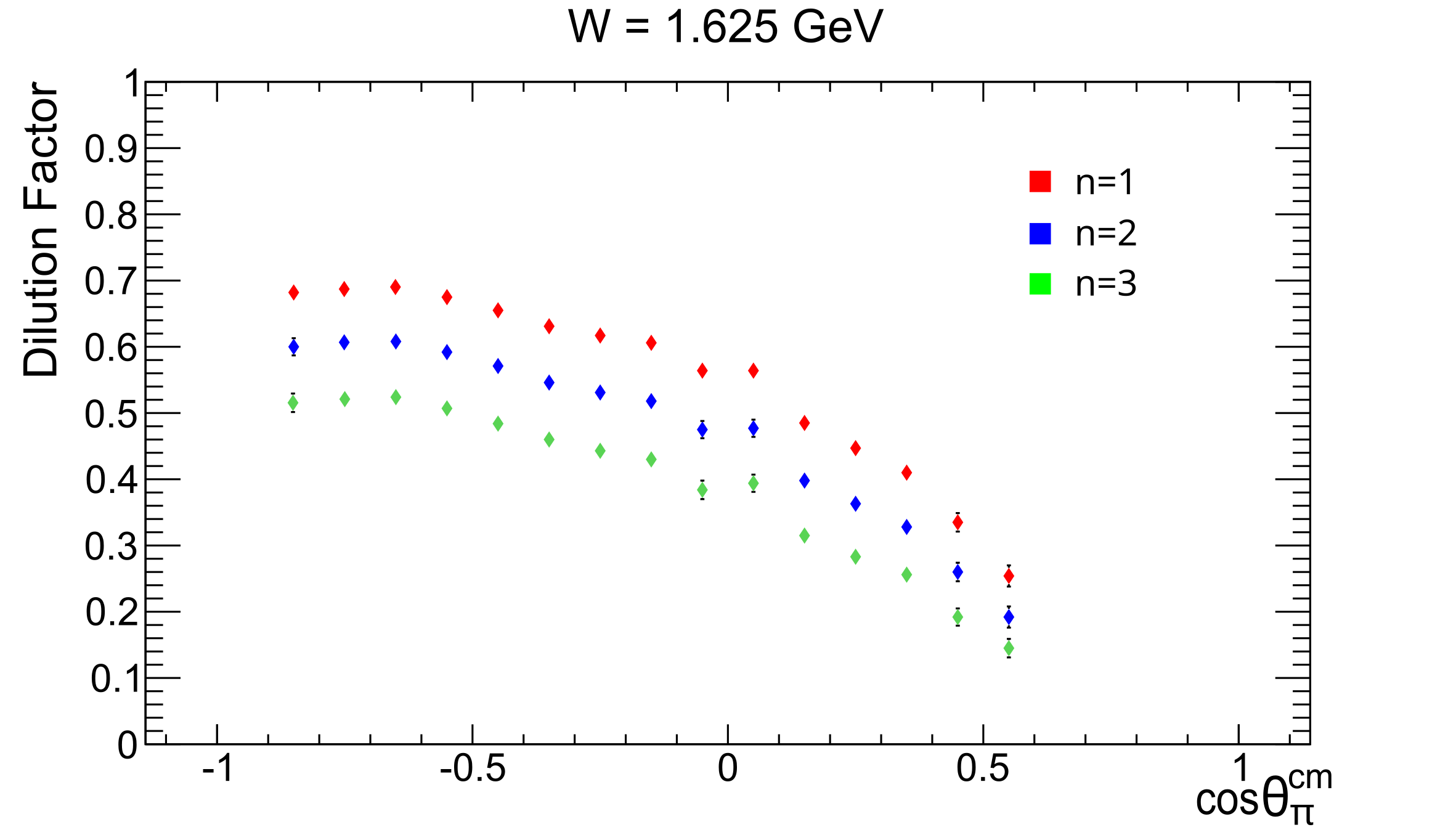}
\includegraphics[width=3.4in]{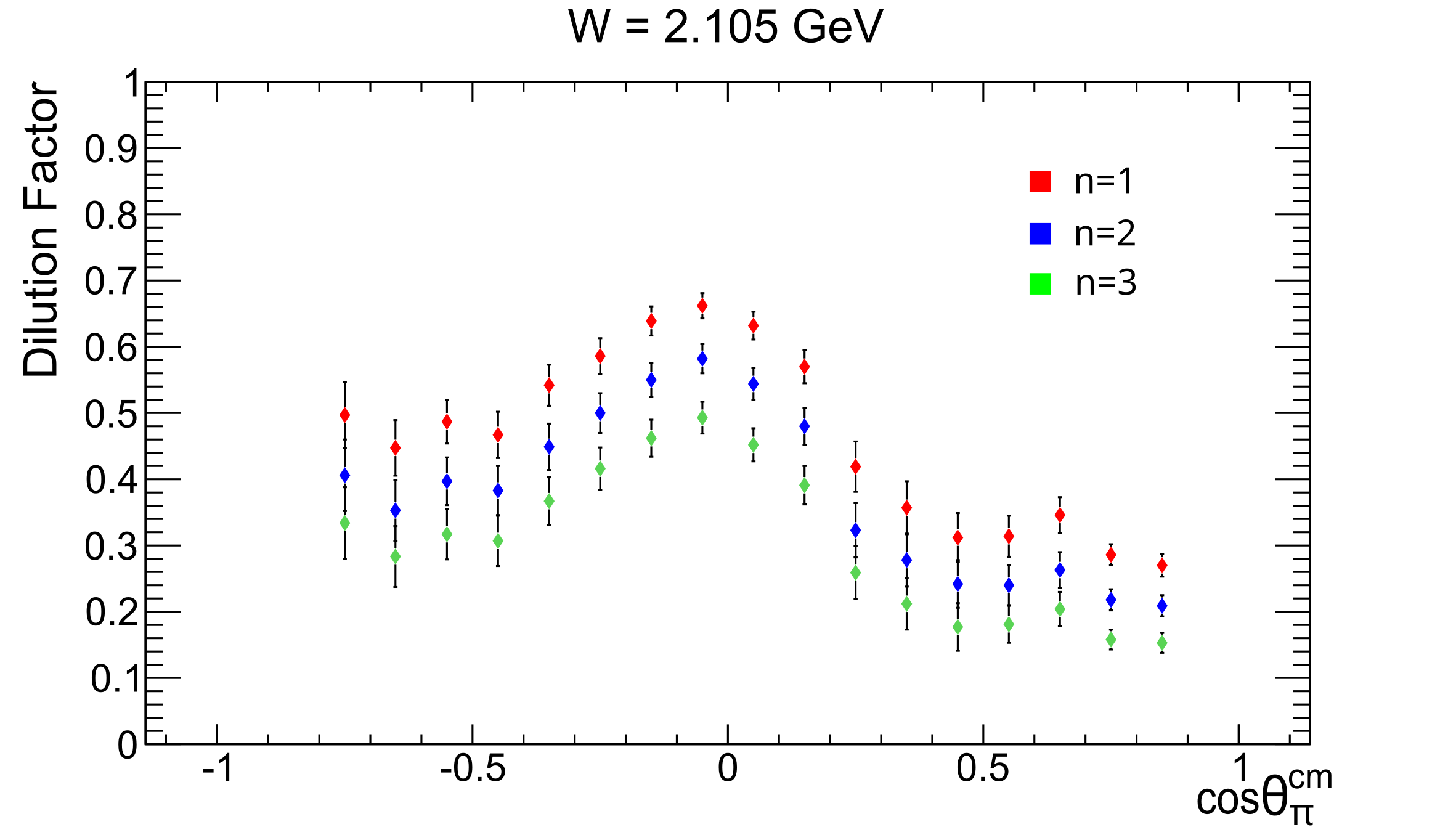}
\caption[Examples of the dilution distributions for various data ranges]{Examples of the dilution-factor distributions for various data-range selections. The selections of $n=1$ are shown in red. The selections of $n=2$ are shown in blue. The selections of $n=3$ are shown in green. The dilution-factor distribution for each width of selection is generally smooth. In this analysis $n = 2$ has been used for the calculation of the results.
\label{fig:dilution}}
\end{figure}

The dilution factor varies as the missing-mass-squared resolution and background contributions change with energy and angle.  For the selection of $n$ in Eq.~(\ref{eq:select}), a balance needs to be found between the statistical uncertainty from the number of events and the magnitude of the dilution factor.  Figure~\ref{fig:dilution} shows the distribution of dilution factors for various data-range selections, $|m^2_X-m^2_0| < n\sigma_H$. The cases of $n=1$, 2, and 3 are shown in three different colors. In all cases, the dilution-factor generally follows a smooth distribution.  As expected, tighter cuts on the free-proton signal lead to increased dilution factors.  In this analysis, $n=2$ has been used for the calculation of the results as this value brings most ($95\%$) events into the analysis and retains a large fraction of the dilution factor relative to the largest possible dilution factor of a given bin.

The further analysis is based on the extraction of the (diluted) polarization observables from the butanol-target data and the correction of those with the dilution factor to obtain the polarization observables for the free proton.  This analysis is only valid if the background is unpolarized and if the dilution factor has been extracted from unpolarized data.  Both of these conditions have been verified in the systematic studies.

\section{Data Analysis}
\subsection{Extraction of Observables}
\label{sec:mom}

The polarization observables $T$ and $F$ for each $(W, \cos\theta_{\pi}^{cm})$ bin were extracted from normalized moments of the selected events, Eq.~(\ref{eq:select}), of that bin. From the expression of the polarized cross section, Eq.~(\ref{eq:polcs}), the integrated normalized polarized yield is
\begin{eqnarray}
Y = &&\frac{1}{2\pi}\int_0^{2\pi}Y_\text{unpol}A(\varphi)\nonumber \\
&&\times (1+P_TT\sin\varphi+P_TP_{\odot} F\cos\varphi)d\varphi,
\end{eqnarray}
where $Y_\text{unpol}$ is proportional to the unpolarized cross section and $A(\varphi)$ is the average acceptance of the detector in the bin of interest. Additionally, the $\sin m\varphi$ and the $\cos m\varphi$ moments are defined as
\begin{eqnarray}
Y_{\sin m\varphi} = &&\frac{1}{2\pi}\int_0^{2\pi}Y_\text{unpol}A(\varphi)(1+P_TT\sin\varphi\nonumber \\
&&+P_TP_{\odot} F\cos\varphi)\sin m\varphi\; d\varphi\quad\text{and}\\
Y_{\cos m\varphi} = &&\frac{1}{2\pi}\int_0^{2\pi}Y_\text{unpol}A(\varphi)(1+P_TT\sin\varphi\nonumber \\
&&+P_TP_{\odot} F\cos\varphi)\cos m\varphi\; d\varphi.
\end{eqnarray}

Experimentally, the normalized moments are obtained for each kinematic bin as
\begin{align}
Y &= \frac{\sum_i(1)}{N^C}, \\
Y_{\sin m\varphi} &= \frac{\sum_i(\sin m\varphi_i)}{N^C},\quad\text{and} \\
Y_{\cos m\varphi} &= \frac{\sum_i(\cos m\varphi_i)}{N^C},
\end{align}
where the sums are taken over the butanol-target events that satisfy the $\gamma p \to \pi^0 p$ kinematics, Eq.~(\ref{eq:select}).  A relative normalization of the sums is given by the number of carbon-target events $N^C$ from the respective setting, which is proportional to the luminosity.  An overall absolute normalization cancels in the final expressions for the observables and can be neglected. Since there are four combinations of two target-polarization directions and two photon-beam helicities, each of the normalized yields or moments can be categorized into four groups. In this analysis $+/-$ stands for the photon-beam helicity and ${\leftarrow}/{\rightarrow}$ for the target-polarization direction. With the dilution factor, the normalized yields or moments of the free-proton data were obtained from those of the butanol data:
\begin{align}
Y^p &= hY \label{eq:dilY}, \\
Y_{\cos 2\varphi}^{p\rightarrow} + Y_{\cos 2\varphi}^{p\leftarrow} &= h(Y_{\cos 2\varphi}^{\rightarrow} + Y_{\cos 2\varphi}^{\leftarrow}).
\label{eq:dilYc2}
\end{align}
Equations (\ref{eq:dilY}) and (\ref{eq:dilYc2}) are only valid if the bound-nucleon background from the butanol target is unpolarized.

The differences of the $\sin \varphi$ and the $\cos \varphi$ moments between different combinations of target or beam polarizations come only from the free-proton events and are not diluted:
\begin{eqnarray}
&&Y_{\sin \varphi}^{p\rightarrow+} + Y_{\sin \varphi}^{p\rightarrow-} - Y_{\sin \varphi}^{p\leftarrow+} - Y_{\sin \varphi}^{p\leftarrow-} \nonumber \\
= &&Y_{\sin \varphi}^{\rightarrow+} + Y_{\sin \varphi}^{\rightarrow-} - Y_{\sin \varphi}^{\leftarrow+} - Y_{\sin \varphi}^{\leftarrow-},\\
&&Y_{\cos \varphi}^{p\rightarrow+} - Y_{\cos \varphi}^{p\rightarrow-} - Y_{\cos \varphi}^{p\leftarrow+} + Y_{\cos \varphi}^{p\leftarrow-} \nonumber \\
= &&Y_{\cos \varphi}^{\rightarrow+} - Y_{\cos \varphi}^{\rightarrow-} - Y_{\cos \varphi}^{\leftarrow+} + Y_{\cos \varphi}^{\leftarrow-}.
\end{eqnarray}

Limited detector acceptances can lead to instrumental asymmetries. Instrumental asymmetries were measured and corrected for with the moments method assuming the acceptance was constant over the run time. The acceptance function $A(\varphi)$ can be expanded in a Fourier series,
\begin{align}
\label{eq:acc}
A(\varphi) = a_0+\sum_{n=1}^{\infty}(a_n\cos n\varphi +b_n\sin n\varphi).
\end{align}

The differences of the normalized moments between different combinations of target or beam polarizations can be expressed as
\begin{align}
Y_{\sin \varphi}^{\rightarrow} - Y_{\sin \varphi}^{\leftarrow} = \frac{1}{2}Y_\text{unpol}(P_T^{\rightarrow}+P_T^{\leftarrow})(a_0-a_2)T
\end{align}
and
\begin{eqnarray}
Y_{\cos \varphi}^{\rightarrow+} - Y_{\cos \varphi}^{\rightarrow-} - Y_{\cos \varphi}^{\leftarrow+} + Y_{\cos \varphi}^{\leftarrow-} = &&\frac{1}{2}Y_\text{unpol}P_{\odot}(P_T^{\rightarrow}+P_T^{\leftarrow})\nonumber \\
&&\times (a_0+a_2)F.
\end{eqnarray}

The terms with Fourier coefficients $Y_\text{unpol}a_0$ and $Y_\text{unpol}a_2$ can be obtained from
\begin{align}
P_T^{\leftarrow}Y^{\rightarrow} + P_T^{\rightarrow}Y^{\leftarrow} = Y_\text{unpol}(P_T^{\rightarrow}+P_T^{\leftarrow})a_0
\end{align}
and
\begin{align}
P_T^{\leftarrow}Y_{\cos 2\varphi}^{\rightarrow} + P_T^{\rightarrow}Y_{\cos 2\varphi}^{\leftarrow} = Y_\text{unpol}(P_T^{\rightarrow}+P_T^{\leftarrow})a_2.
\end{align}

By utilizing different combinations of target or beam polarizations, after determining the acceptance effect, the polarization observables $T$ and $F$ can be obtained as
\begin{equation}
T = \frac{1}{h}\frac{2(Y_{\sin \varphi}^{\rightarrow} - Y_{\sin \varphi}^{\leftarrow})}{P_T^{\leftarrow}(Y^{\rightarrow}-Y_{\cos 2\varphi}^{\rightarrow}) + P_T^{\rightarrow}(Y^{\leftarrow}-Y_{\cos 2\varphi}^{\leftarrow})}
\end{equation}
and
\begin{equation}
F = \frac{1}{h}\frac{2(Y_{\cos \varphi}^{\rightarrow+} - Y_{\cos \varphi}^{\rightarrow-} - Y_{\cos \varphi}^{\leftarrow+} + Y_{\cos \varphi}^{\leftarrow-})}{P_{\odot}P_T^{\leftarrow}(Y^{\rightarrow}+Y_{\cos 2\varphi}^{\rightarrow}) + P_{\odot}P_T^{\rightarrow}(Y^{\leftarrow}+Y_{\cos 2\varphi}^{\leftarrow})}.
\label{eq:F}
\end{equation}

Both of the expressions for the observables have the form of $\frac{u}{hv}$. The variance of $\frac{u}{hv}$ can be expressed as
\begin{align}
\text{Var}\left(\frac{u}{hv}\right) =
  \frac{u^2}{h^2v^2}\left(\frac{\sigma^2_h}{h^2}+\frac{\sigma^2_u}{u^2}+\frac{\sigma^2_v}{v^2}-\frac{\text{Cov}(u,v)}{uv}\right), \label{eq:var}
\end{align}
where $\sigma^2_h$, $\sigma^2_u$, and $\sigma^2_v$ are the variances of $h$, $u$, and $v$, respectively, and $\text{Cov}(u,v)$ is the covariance between $u$ and $v$; the dilution factor $h$ is uncorrelated.  This relation was used to calculate the statistical uncertainties of the polarization observables $T$ and $F$, respectively.

For the statistical uncertainty of the observable $T$ this means
\begin{align}
u &= 2(Y_{\sin \varphi}^{\rightarrow} - Y_{\sin \varphi}^{\leftarrow})
    \quad \text{and}\\
v &= P_T^{\leftarrow}(Y^{\rightarrow}-Y_{\cos 2\varphi}^{\rightarrow}) + P_T^{\rightarrow}(Y^{\leftarrow}-Y_{\cos 2\varphi}^{\leftarrow}).
\end{align}
It follows then, that
\begin{eqnarray}
\sigma^2_u = &&\frac{2}{N^{\rightarrow}}(Y^{\rightarrow}-Y_{\cos 2\varphi}^{\rightarrow}) + \frac{2}{N^{\leftarrow}}(Y^{\leftarrow}-Y_{\cos 2\varphi}^{\leftarrow}),\\
\sigma^2_v = &&\frac{P_T^{\leftarrow2}}{2N^{\rightarrow}}(3Y^{\rightarrow}-4Y_{\cos 2\varphi}^{\rightarrow}+Y_{\cos 4\varphi}^{\rightarrow})\nonumber \\
&&+ \frac{P_T^{\rightarrow2}}{2N^{\leftarrow}}(3Y^{\leftarrow}-4Y_{\cos 2\varphi}^{\leftarrow}+Y_{\cos 4\varphi}^{\leftarrow}),
\end{eqnarray}
and
\begin{eqnarray}
\text{Cov}(u,v) = &&\frac{P_T^{\leftarrow}}{N^{\rightarrow}}(3Y_{\sin \varphi}^{\rightarrow}-Y_{\sin 3\varphi}^{\rightarrow})\nonumber \\
&&- \frac{P_T^{\rightarrow}}{N^{\leftarrow}}(3Y_{\sin \varphi}^{\leftarrow}-Y_{\sin 3\varphi}^{\leftarrow}).
\end{eqnarray}
For the statistical uncertainty of the observable $F$, we find
\begin{align}
u &= 2(Y_{\cos \varphi}^{\rightarrow+} - Y_{\cos \varphi}^{\rightarrow-} - Y_{\cos \varphi}^{\leftarrow+} + Y_{\cos \varphi}^{\leftarrow-})\quad\text{and}\\
v &= P_{\odot}P_T^{\leftarrow}(Y^{\rightarrow}+Y_{\cos 2\varphi}^{\rightarrow}) + P_{\odot}P_T^{\rightarrow}(Y^{\leftarrow}+Y_{\cos 2\varphi}^{\leftarrow}),
\end{align}
and consequently
\begin{eqnarray}
\sigma^2_u = &&\frac{2}{N^{\rightarrow}}(Y^{\rightarrow}+Y_{\cos 2\varphi}^{\rightarrow}) + \frac{2}{N^{\leftarrow}}(Y^{\leftarrow}+Y_{\cos 2\varphi}^{\leftarrow}),\\
\sigma^2_v = &&\frac{P_{\odot}^2P_T^{\leftarrow2}}{2N^{\rightarrow}}(3Y^{\rightarrow}+4Y_{\cos 2\varphi}^{\rightarrow}+Y_{\cos 4\varphi}^{\rightarrow}) \nonumber \\
&&+ \frac{P_{\odot}^2P_T^{\rightarrow2}}{2N^{\leftarrow}}(3Y^{\leftarrow}+4Y_{\cos 2\varphi}^{\leftarrow}+Y_{\cos 4\varphi}^{\leftarrow}),
\end{eqnarray}
\begin{multline}
\text{Cov}(u,v) = \frac{P_{\odot}P_T^{\leftarrow}}{N^{\rightarrow}}(3Y_{\cos \varphi}^{\rightarrow+}-3Y_{\cos \varphi}^{\rightarrow-}+Y_{\cos 3\varphi}^{\rightarrow+}-Y_{\cos 3\varphi}^{\rightarrow-}) \\ - \frac{P_{\odot}P_T^{\rightarrow}}{N^{\leftarrow}}(3Y_{\cos \varphi}^{\leftarrow+}-3Y_{\cos \varphi}^{\leftarrow-}+Y_{\cos 3\varphi}^{\leftarrow+}-Y_{\cos 3\varphi}^{\leftarrow-}).
\end{multline}

\subsection{Target-Polarization Orientation}

Using Eq.~(\ref{eq:phi}), the polarized cross section can be expressed as
\begin{eqnarray}
 \frac{d\sigma}{d\Omega} = &&\frac{d\sigma}{d\Omega}_\text{unpol}[1+ P_T T \sin(\varphi_0-\varphi_\pi^\text{lab}) \nonumber \\
&&+ P_T P_{\odot} F \cos(\varphi_0-\varphi_\pi^\text{lab})].
\label{eq:polcslab}
\end{eqnarray}
Butanol-data were used to determine the target-polarization orientation $\varphi_0$ in the lab frame. To ensure that the results are not affected by the acceptance changes among run-groups, the helicity-dependent part of the polarized yield was utilized to calculate $\varphi_0$, based on the term $P_T P_{\odot} F \cos(\varphi_0-\varphi_\pi^\text{lab})$. The integrated normalized polarized yield can be expressed as

\begin{eqnarray}
  Y = &&\frac{1}{2\pi}\int_0^{2\pi}Y_\text{unpol}A(\varphi_\pi^\text{lab})[1 + P_TT(\sin\varphi_0\cos\varphi_\pi^\text{lab} \nonumber \\
&&-\cos\varphi_0\sin\varphi_\pi^\text{lab}) + P_TP_{\odot} F(\cos\varphi_0\cos\varphi_\pi^\text{lab}\nonumber \\
&&+\sin\varphi_0\sin\varphi_\pi^\text{lab})] d\varphi_\pi^\text{lab},
  \label{eq:Yphi0}
\end{eqnarray}
where $A(\varphi_\pi^\text{lab})$ is the average acceptance in the lab frame.  Similarly to the extraction of the polarization observable $F$, the moments method has been used to determine $F\sin\varphi_0$, which enters in Eq.~(\ref{eq:Yphi0}) through the helicity-dependent $\sin\varphi^\text{lab}_\pi$ term, and  $F\cos\varphi_0$, which enters through the helicity-dependent $\cos\varphi^\text{lab}_\pi$ term.  The ratio of these terms give $\tan\varphi_0$ and is expressed in terms of the moments:

\begin{widetext}
\begin{equation}
\label{eq:tanphi0}
\tan \varphi_0 = \frac{(P_T^{\rightarrow}Y_{\sin 2\varphi_\pi^\text{lab}}^{\leftarrow}\!+\!P_T^{\leftarrow}Y_{\sin 2\varphi_\pi^\text{lab}}^{\rightarrow})r \!-\! P_T^{\rightarrow}(Y^{\leftarrow}\!+\!Y_{\cos 2\varphi_\pi^\text{lab}}^{\leftarrow})q\!-\!P_T^{\leftarrow}(Y^{\rightarrow}\!+\!Y_{\cos 2\varphi_\pi^\text{lab}}^{\rightarrow})q}{(P_T^{\rightarrow}Y_{\sin 2\varphi_\pi^\text{lab}}^{\leftarrow}\!+\!P_T^{\leftarrow}Y_{\sin 2\varphi_\pi^\text{lab}}^{\rightarrow})q \!-\! P_T^{\rightarrow}(Y^{\leftarrow}\!-\!Y_{\cos 2\varphi_\pi^\text{lab}}^{\leftarrow})r\!-\!P_T^{\leftarrow}(Y^{\rightarrow}\!-\!Y_{\cos 2\varphi_\pi^\text{lab}}^{\rightarrow})r},
\end{equation}
\end{widetext}
where
\begin{align}
q &= Y_{\sin \varphi_\pi^\text{lab}}^{\rightarrow+} - Y_{\sin \varphi_\pi^\text{lab}}^{\rightarrow-} - Y_{\sin \varphi_\pi^\text{lab}}^{\leftarrow+} + Y_{\sin \varphi_\pi^\text{lab}}^{\leftarrow-}\quad\text{and} \\
r &=Y_{\cos \varphi_\pi^\text{lab}}^{\rightarrow+} - Y_{\cos \varphi_\pi^\text{lab}}^{\rightarrow-} - Y_{\cos \varphi_\pi^\text{lab}}^{\leftarrow+} + Y_{\cos \varphi_\pi^\text{lab}}^{\leftarrow-}.
\end{align}

In the evaluation of the moments, butanol data were used for all kinematic bins within the range of $-0.02~(\text{GeV/c$^2$})^2 < m_X^2 < 0.06~(\text{GeV/c$^2$})^2$ in the missing-mass-squared distribution.  Equation~(\ref{eq:tanphi0}) has two solutions for $\varphi_0$ between $0^\circ$ and $360^\circ$: $116.4^{\circ}\pm1.4^{\circ}$ and $296.3^{\circ}\pm1.4^{\circ}$.  Both values give in the analysis the same magnitudes of the polarization observables $T$ and $F$, but opposite signs.  A comparison of the results with the world dataset shows that $\varphi_0 = 116.4^{\circ}\pm1.4^{\circ}$ is the correct solution.  Equation~(\ref{eq:tanphi0}) takes into account differences in the target polarization, instrumental asymmetries, and acceptance effects, and has been used in the analysis.  However, the simplified expression, 
\begin{align*}
  \tan \varphi_0 \approx \frac{q}{r} = \frac{Y_{\sin \varphi_\pi^\text{lab}}^{\rightarrow+} - Y_{\sin \varphi_\pi^\text{lab}}^{\rightarrow-} - Y_{\sin \varphi_\pi^\text{lab}}^{\leftarrow+} + Y_{\sin \varphi_\pi^\text{lab}}^{\leftarrow-}}{
  Y_{\cos \varphi_\pi^\text{lab}}^{\rightarrow+} - Y_{\cos \varphi_\pi^\text{lab}}^{\rightarrow-} - Y_{\cos \varphi_\pi^\text{lab}}^{\leftarrow+} + Y_{\cos \varphi_\pi^\text{lab}}^{\leftarrow-}},
\end{align*}
makes the main idea more transparent.

The statistical uncertainty of $\varphi_0$, $\delta = 1.4^\circ$, was determined from the propagation of the statistical uncertainties of the moments in the expression Eq.~(\ref{eq:tanphi0}).
It contributes to the uncertainties of the polarization observables $T$ and $F$ through \begin{align}
  T\sin(\varphi\pm\delta) &= T\left(\sin\varphi\cos\delta \pm \cos\varphi\sin\delta\right)\quad\text{and} \\
  F\cos(\varphi\pm\delta) &= F\left(\cos\varphi\cos\delta \mp \sin\varphi\sin\delta\right).
\end{align}

The observable $T$ enters as $T\sin\varphi$ in the polarized cross section, and there is no helicity-independent observable connected to $\cos\varphi$. Similarly, the observable $F$ enters as $F\cos\varphi$ in the polarized cross section, and there is no helicity-dependent observable connected to $\sin\varphi$. The net effect of a non-zero $\delta$ is, therefore, in either case, a dilution of the true value of the observables in the reconstruction by a factor of $\cos\delta > 0.999$ that is negligible.

\section{Systematic Uncertainties}

The systematic uncertainties are mainly from beam and target polarizations and the background. The average systematic uncertainties (relative $\pm$ absolute) are 4.6\% $\pm$ 0.009 for $T$ and 5.5\% $\pm$ 0.007 for $F$. Besides the systematic uncertainties from the degree of beam polarization, the degree of target polarization, the proton misidentification, and the target-polarization orientation mentioned previously, the systematic uncertainties are also from the beam-charge asymmetry~\cite{moller} and several studies. A summary of the systematic uncertainties is given in Table~\ref{tab:sys}.

\begin{table*}[!htb]
\caption[Systematic uncertainties for $T$ and $F$]{Systematic uncertainties for $T$ and $F$.}
\label{tab:sys}
\begin{ruledtabular}
\begin{tabular}{lccccl}
Item & $\sigma(T)$ & $\sigma(F)$ \\
\hline
Beam-charge asymmetry & --- & 0.2\% \\
Degree of beam polarization & --- & 3\% \\
Degree of target polarization & 4\% & 4\% \\
Target-polarization orientation $\varphi_0$ & $< 0.001$ & $< 0.001$ \\
Accidental background & $3.7\%\left(\frac{W}{\text{GeV}} - 1.32\right)$ & $3.7\%\left(\frac{W}{\text{GeV}} - 1.32\right)$ \\
Proton misidentification & $<1$\% & $<1$\% \\
Polarized yield in the dilution factor extraction & $\lesssim 0.2\%$ & --- \\
Background polarization & $\approx 0$ & $\approx 0$ \\
Background subtraction & $\pm 0.006$ & $\pm 0.007$ \\
Run-group acceptance changes & $\pm$0.007 & --- \\
\hline
Total (Avg.) & 4.6\% $\pm$ 0.009  & 5.5\% $\pm$ 0.007 \\
\end{tabular}
\end{ruledtabular}
\end{table*}

\subsection{Accidental Background}
\label{sec:accident}

In the presence of unsubtracted accidental background, the extracted polarization observable, $O^\text{exp}$, where  $O$ stands for $ T$ or $F$, is the yield-weighted average of the polarization observable of the true signal, $O$, and that of the accidental background, $O^\text{acc}$:
\begin{equation}
  O^\text{exp} = \frac{N^\text{true}O +
    N^\text{acc}O^\text{acc}}{N^\text{true}+N^\text{acc}}.
  \end{equation}
The signal-polarization observable of interest can be estimated from the extracted one,
\begin{equation}
  O = \left( 1 + \frac{N^\text{acc}}{N^\text{true}}\right)O^\text{exp} - \frac{N^\text{acc}}{N^\text{true}}O^\text{acc},
  \label{eq:Oacc}
  \end{equation}
and the fraction of accidental background, $N^\text{acc}/N^\text{true}$, if the polarization observable of the background, $O^\text{acc}$, is known.  The effect of this background on the extracted polarization observable has two contributions:  one is a dilution of the true observable and the other a bias of the experimental value that depends on the polarization of the background.  
The peaks at multiples of approximately $\pm 2~\text{ns}$ in the coincidence-time distribution are from the detection of uncorrelated photon and proton events.  Assuming that a similar random background exists also beneath the timing peak of the true coincidences at $\approx 0~\text{ns}$, the fraction of accidental background can be estimated. It varies from about 1\% to 4\%. The $W$ dependence is nearly linear and can be expressed as:
\begin{equation}
  \frac{N^\text{acc}}{N^\text{true}} = 3.7\%\left(\frac{W}{\text{GeV}} - 1.32\right).
  \label{eq:wd}
  \end{equation}

The statistics in this experiment is not sufficient to study the value of $O^\text{acc}$ in detail.  The worst but very unrealistic case of $O^\text{acc} = -O^\text{exp}/|O^\text{exp}|$ would lead to uncertainties of up to $4\% O + 0.04$ at the highest values of $W$. The values of the extracted polarization observables are $\bar T = 0.35$ and $\bar F = -0.21$, respectively, when averaged over all energies and angles.  If $O^\text{acc} \approx \bar F, \bar T$, the effect is mostly a dilution of up to 4\%, which is on average compensated by the asymmetry of the background --- also for observable $F$, as photons of neighboring beam buckets have almost always the same helicity.  Chances are, however, that the photon energy of the random photon is biased towards the photon energy of the true events due to the cut in the missing mass. If then $O^\text{acc} \approx O^\text{exp}$, the two effects of the accidental background would approximately cancel in each kinematic bin. We assume, conservatively, $O^\text{acc} \approx 0$. The accidental background then leads to a relative error that is equal to the $W$-dependent dilution $N^\text{acc}/N^\text{true}$. We have included this effect of the accidental background as $W$-dependent systematic uncertainty in the error budget.

\subsection{Dilution Factor from Unpolarized Yields}
\label{sec:unpolY}

To be viable for the analysis, the dilution factor has to be extracted from unpolarized yields.  As stated above, the sums in Eq.~(\ref{eq:chi}) are over both helicity states, all run-groups, and all azimuthal angles.  The summation over the photon-helicity states removes the effect of the polarization observable $F$ to the cross section entirely.  The summation over all run-groups does not fully remove the effect of $T$ as the data for the two target-polarization orientations were not taken with the same statistics.  The asymmetry in the statistics as measured by carbon-target yields is $A_Y \approx 0.06$ for the combined run-groups one through five; see Table~\ref{tab:g9b}.  Also, the integration over all azimuthal angles does not completely render an unpolarized yield due to angular-dependent detector inefficiencies. The relevant instrumental asymmetry is of the order $A_\phi \approx 0.08$.  The limited target polarization reduces the polarized-yield asymmetry further: $A_T \approx 0.80$.  Finally, the polarization observable $T$ is typically smaller than 0.5; $T \approx 0.5$. Combined, these effects could cause a deviation of the free-proton-signal yield, Eq.~(\ref{eq:funcS}), from its unpolarized value by a relative fraction of $ A_Y A_\phi A_T T \lesssim 0.002$. In the presence of such an asymmetry, the dilution factor would be off by the same relative value.  This analysis has not attempted to correct the missing-mass-squared distribution for the effect of the residual asymmetry as this effect is much smaller than other contributions to the systematic uncertainties related to the background correction with dilution factors.

\subsection{Event Selection}
\label{sec:backY}

As expected and shown in Fig.~\ref{fig:dilution}, the dilution factor varies with the missing-mass-squared range that is used in the analysis, Eq.~(\ref{eq:select}).  The final results of the extracted polarization observables, however, need to be independent of the particular choice of the range.  To study possible systematic effects, comparisons of the differences between observables from ranges of selection with $n = 1$, 2, and 3 were made. The absolute values of the mean of Gaussian fits to the distribution of the differences between observables under two consecutive data-range selections act as indicators of possible systematic effects. For the observables $T$ and $F$, the systematic uncertainties are estimated based on the larger of the obtained mean differences, 0.006 and 0.007, respectively.  However, it is likely that statistical uncertainties dominate these values, and the systematic differences are smaller.  In fact, the absolute values of the mean values are closer to zero for the comparison of the higher-statistics sets, $\pm2\sigma_H$ and $\pm3\sigma_H$.

\subsection{Background}

There are two background contributions to the polarized free-proton $\gamma p \to \pi^0 p$ reaction: background from other reaction channels off the polarized free-protons, the most relevant of which is the double-pion photoproduction and background from reactions off bound nucleons.  Both of these background contributions can be polarized or unpolarized.

Background from unpolarized reactions off bound-nucleons is accounted for in this analysis through the correction of the extracted raw asymmetries with a dilution factor.  Except for the free protons, all nucleons in the target material are unpolarized. There is no polarization asymmetry in the cross section for the $\gamma p \to \pi^0 p$ reaction with a circularly polarized beam off unpolarized protons.  Such polarization asymmetries are possible for three-body final states, like double-pion photoproduction.  These asymmetries cancel in the polarized-proton yield and moment asymmetries that take differences for opposite target-polarization orientations.

Possible background from free-proton reactions was studied in detail.  A polarized background would bias the extracted results, and unpolarized background would dilute the results in a way that is not corrected for by the dilution factor. In one analysis, raw target ($T$) and beam-target ($F$) asymmetries were determined for data in various ranges of the missing-mass-squared distribution.  Relative to the $m^2_{\pi^0}$ peak in the distribution and in units of the peak's width, these regions are: $[-6, -3], [-2, 2], [3, 4], [4, 5]$, and $[5, 6]$. It appears unlikely that tails from that polarized background leak under the central peak. In a second study the final results for the observables $T$ and $F$ were compared for each kinematic bin of this analysis for data from the regions $[-2,0]$ and $[0,2]$, respectively.  Free-proton background for the final state $pX$ will have missing masses $m_X > m_{\pi^0}$.  If such background is present, it is expected to affect the results for the polarization observables more strongly in the missing-mass-squared range $[0,2]$ compared to the range $[-2,0]$. There is no indication that the results for the high-missing-mass results are systematically different from the results at low-missing-mass.  

\subsection{Detector Acceptance}
\label{sec:acc}

The average detector acceptance changed over the run time for the runs used in the analysis. As discussed before, the average acceptance was assumed to remain constant between run-groups in the moment-method analysis. In the case of the observable $F$, this condition was automatically fulfilled as the beam helicity flipped rapidly to ensure the same average acceptance of events from both helicity states. In the case of the observable $T$, since the target polarization was flipped run-group by run-group, the acceptance was not guaranteed to remain the same for all run-groups. In fact, in a further study, we have determined the Fourier coefficients of the acceptance, Eq.~(\ref{eq:acc}), using carbon-target events.  As these events are unpolarized, any non-zero value of $b_1$ indicates an instrumental asymmetry due to a limited detector acceptance, which would affect the extraction of the observable $T$. The instrumental asymmetry could be due to changes in the average acceptance caused by detector problems or by effects associated with the different holding-field orientations. The study showed that the coefficient $b_1/a_0$ changed significantly between run-groups 3 and 4, but remained approximately constant otherwise. To account for that change in $A(\varphi)$, the whole data set was divided into two parts, run-groups 1 to 3 and 4 to 5, respectively, to ensure the constancy of the acceptance in each part. Results of the observable $T$ were extracted from these two parts separately. The two results were found to be consistent. The final results were taken as the weighted average of the two. For the remaining small differences in the acceptance within each part, the Fourier coefficients indicate a 0.007 systematic uncertainty for $T$ estimated as the standard deviation of the raw asymmetry $b_1/a_0$ in each of the two parts divided by the average target polarization and dilution factors.

\section{Results}

\begin{figure}[!htb]
  \centering
\includegraphics[width=3.4in]{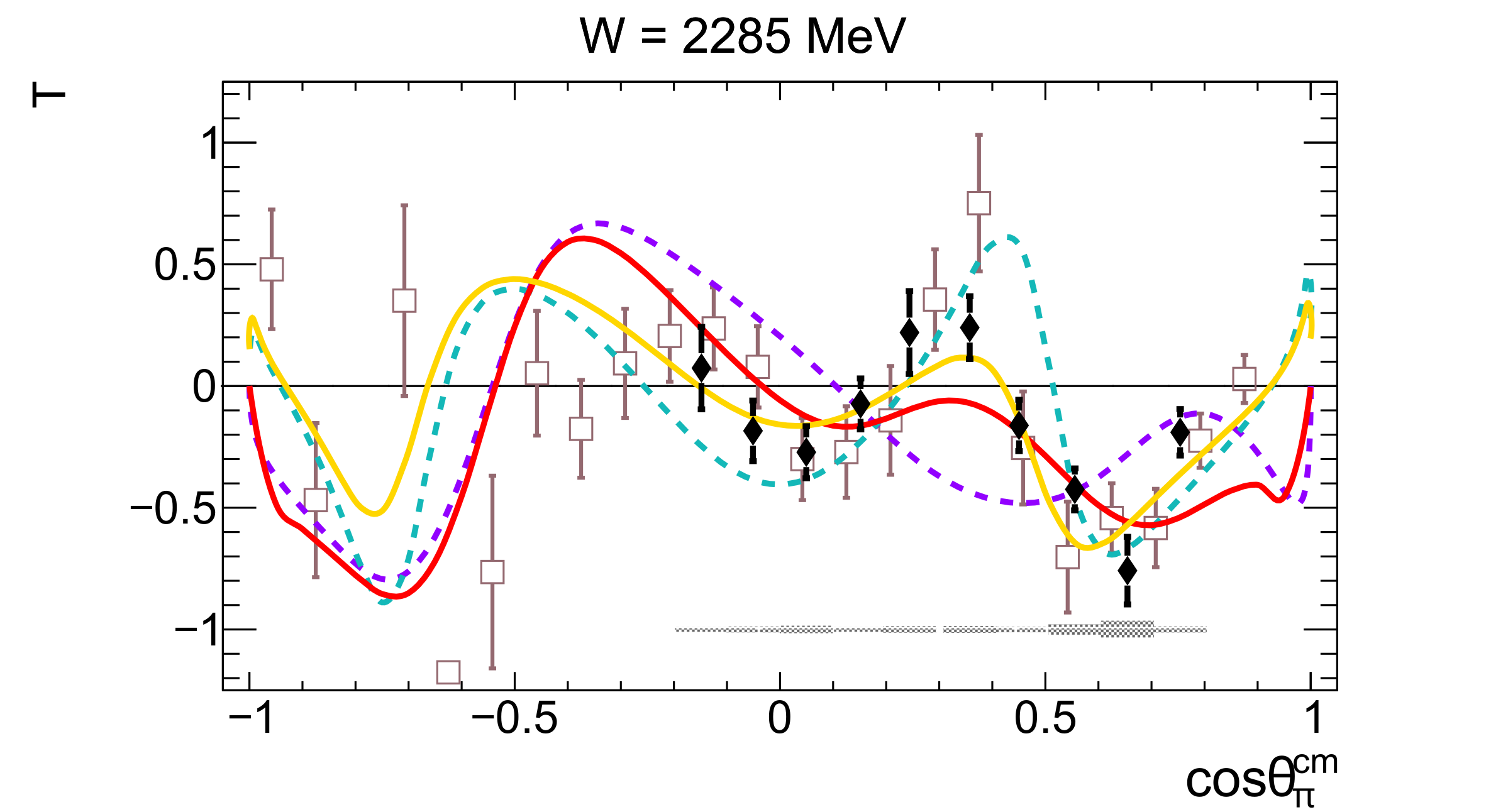}
\includegraphics[width=3.4in]{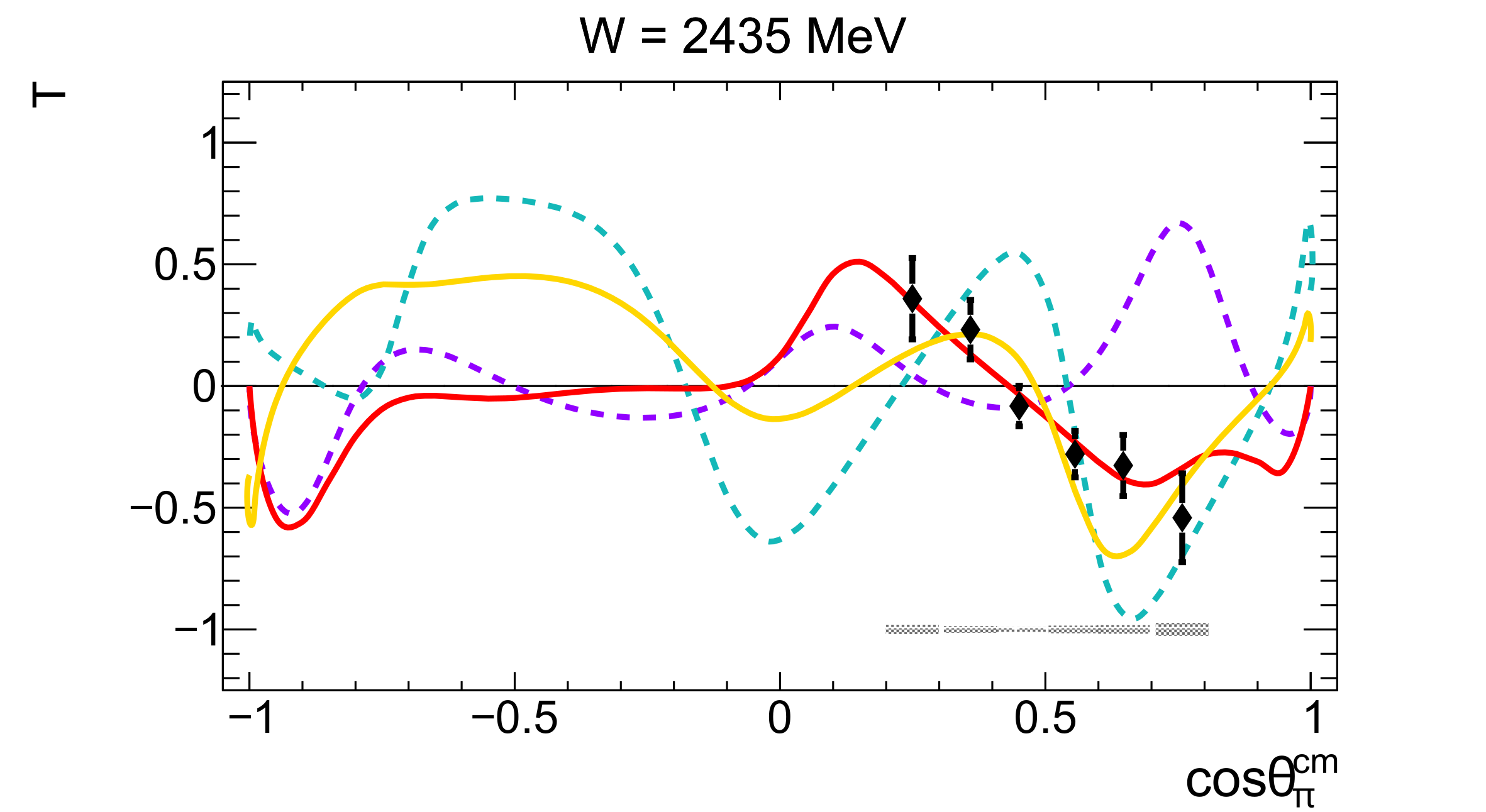}
\includegraphics[width=3.4in]{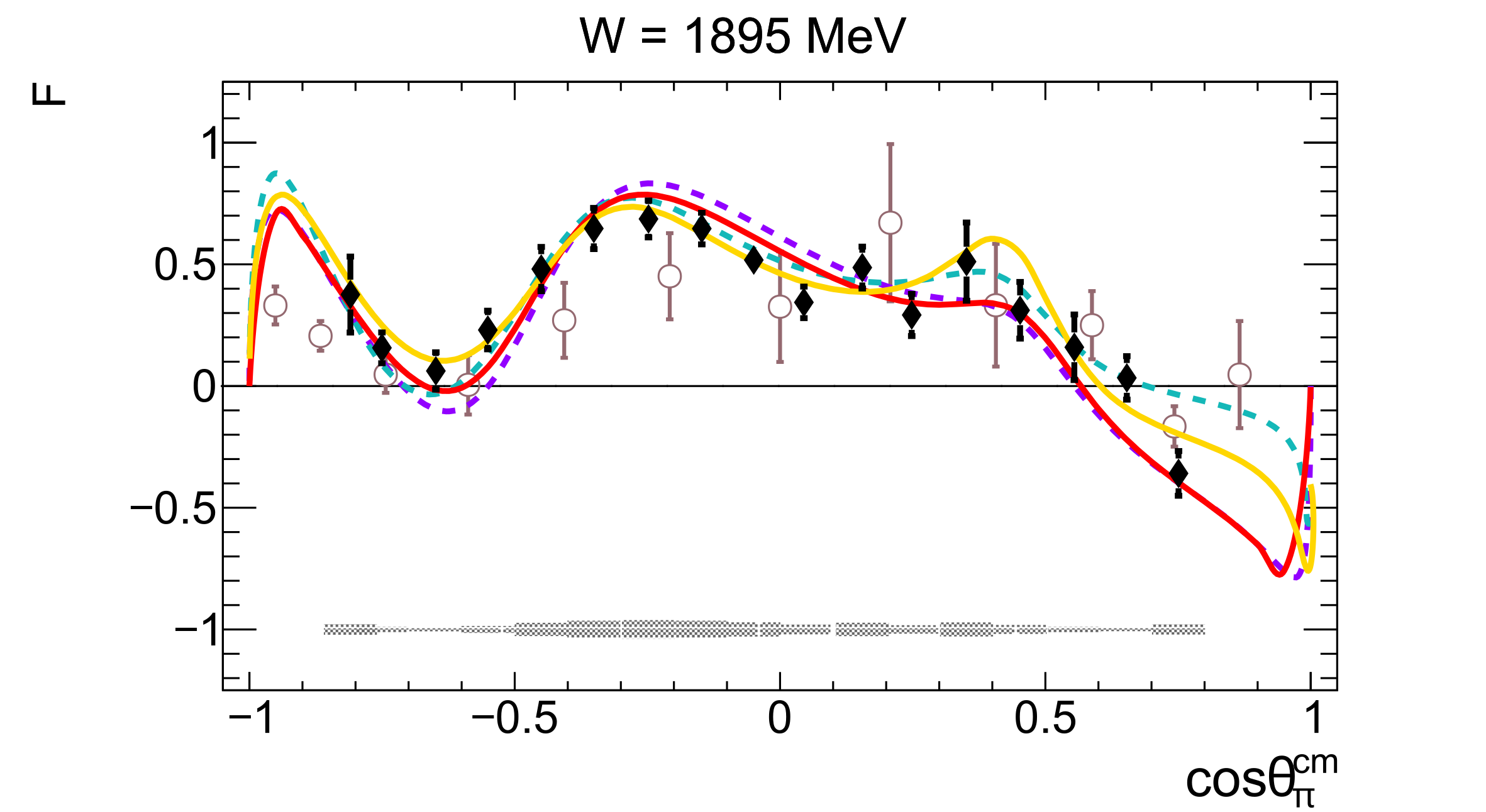}
\includegraphics[width=3.4in]{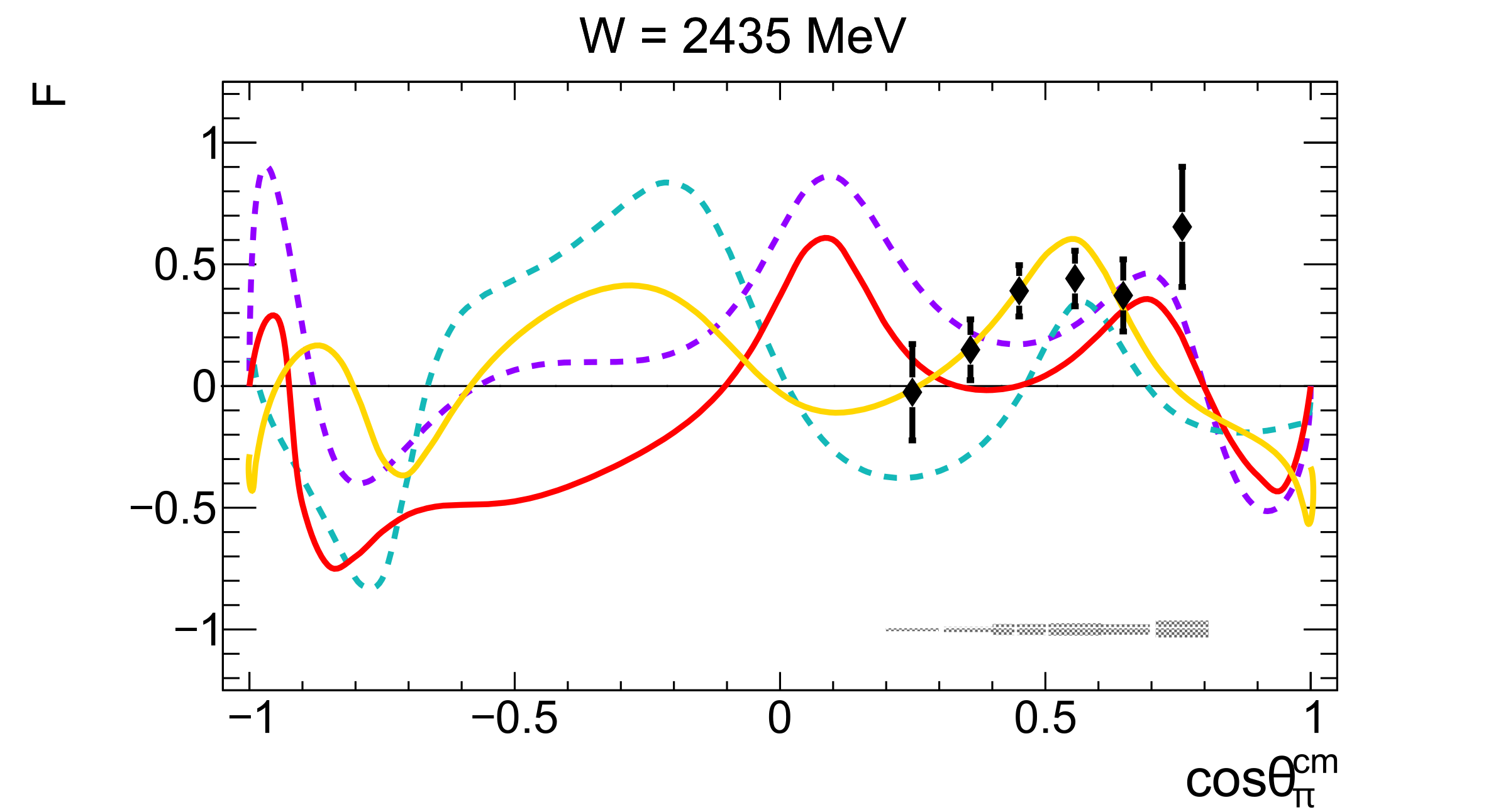}
\caption[]{Examples of the observables $T$ and $F$. The results of this analysis are shown in black. The gray band indicates the size of the systematic uncertainties. The SAID (KX23) (with the present result included) and SAID (SM22)~\cite{sm22said} solutions are shown in red and purple, respectively. The J{\"u}Bo (2023) (with the present result included) and J{\"u}Bo (2022)~\cite{Jubo} solutions are shown in yellow and teal, respectively. Data from MAMI~\cite{AN} are shown as circles and data from CBELSA/TAPS~\cite{HA2} are shown as squares.
\label{fig:kx23TF}}
\end{figure}

The results of polarization observables are available in the center-of-mass energy $W$ ranges from 1445 MeV to 2525 MeV. There are 488 data points in bins that are 30 MeV wide in $W$ and 0.1 wide in $\cos\theta_{\pi}^{cm}$. The results are compared with solutions from SAID (KX23) (with the present result included), J{\"u}Bo (2023) (with the present result included), BnGa (2019)~\cite{CBELSATAPS:2019ylw}, and MAID (2007)~\cite{MAID}. The present SAID, J{\"u}Bo, BnGa, and MAID model predictions generally agree with the data in the energy range overlapped with previous measurements from MAMI~\cite{AN} and CBELSA/TAPS~\cite{HA2} but significant deviations are observed, mostly in the range of $W$ above 2100 MeV. Examples of the fits with the present data included and predictions without the present data included are shown in Fig.~\ref{fig:kx23TF}.

\begin{figure*}
\centering
\includegraphics[width=7.1in]{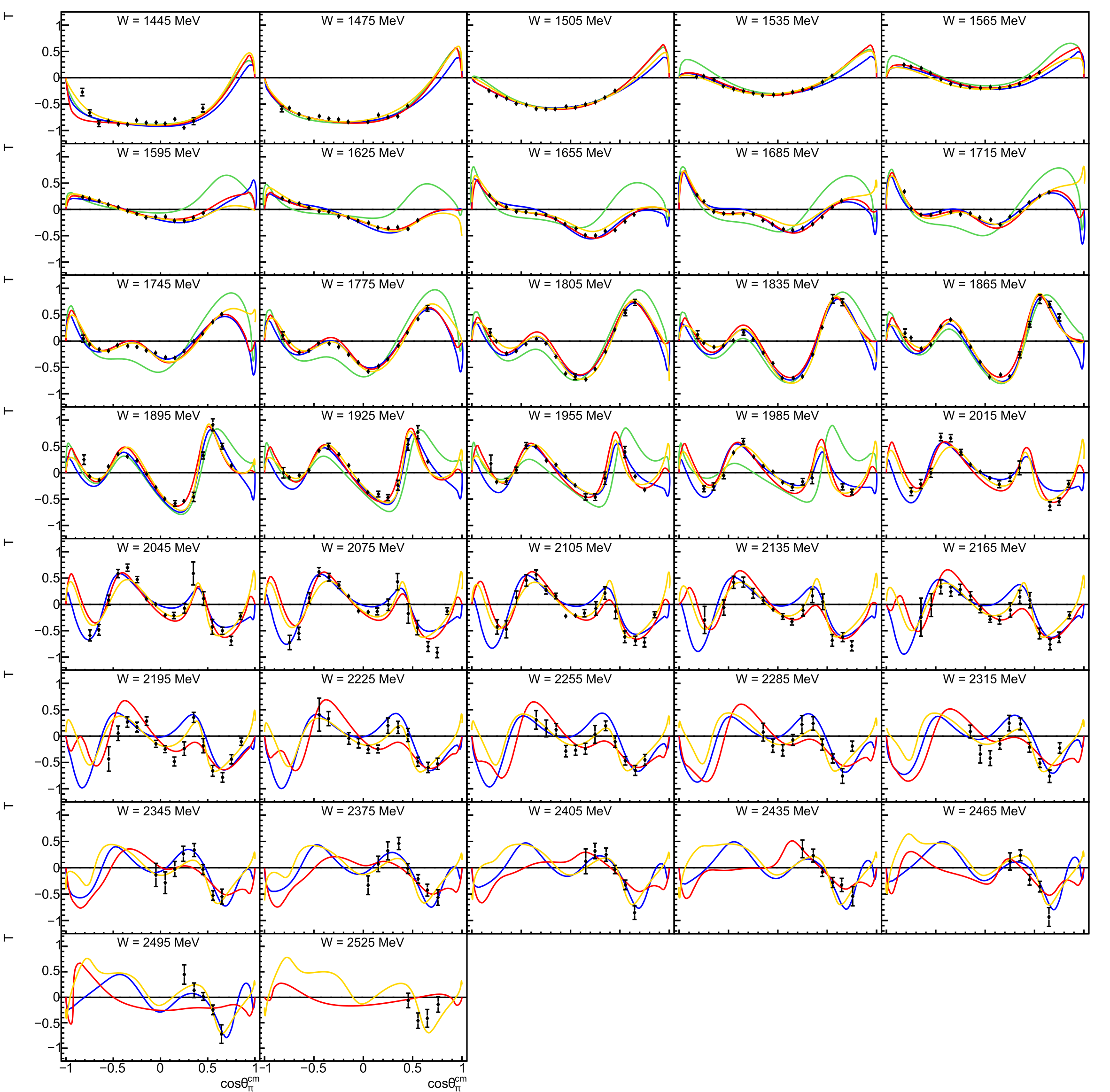}
\caption[T]{Polarization observable $T$ in the $\gamma p \to \pi^0 p$ reaction as a function of $\cos\theta_{\pi}^{cm}$ for all available bins of the center-of-mass energy $W$. The results of this analysis are shown in black. The SAID (KX23), J{\"u}Bo (2023), BnGa (2019)~\cite{CBELSATAPS:2019ylw}, and MAID (2007)~\cite{MAID} solutions are shown in red, gold, blue, and green, respectively.
\label{fig:obrT}}
\end{figure*}

\begin{figure*}
\centering
\includegraphics[width=7.1in]{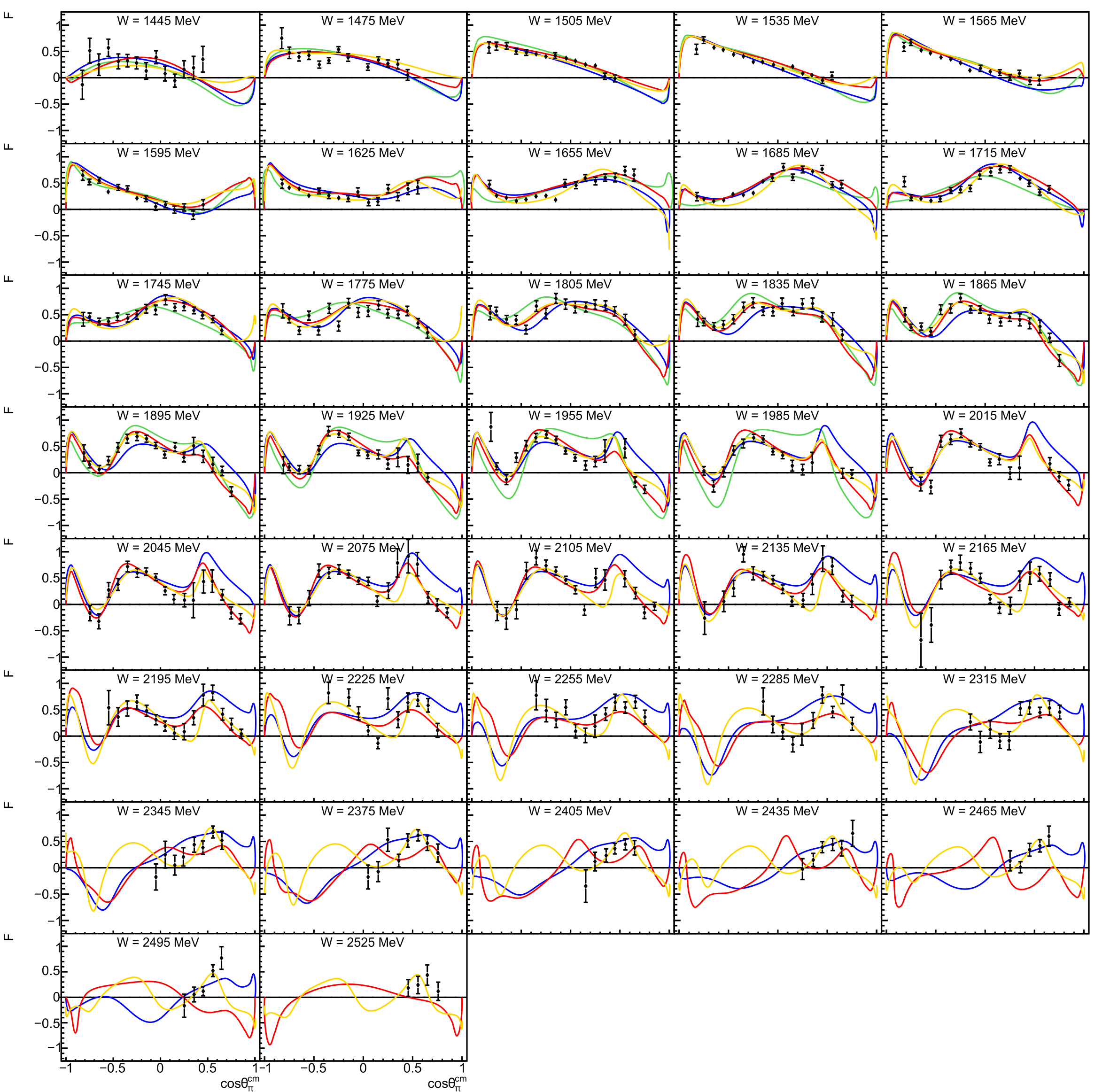}
\caption[F]{Polarization observable $F$ in the $\gamma p \to \pi^0 p$ reaction as a function of $\cos\theta_{\pi}^{cm}$ for all available bins of the center-of-mass energy $W$. The results of this analysis are shown in black. The SAID (KX23), J{\"u}Bo (2023), BnGa (2019)~\cite{CBELSATAPS:2019ylw}, and MAID (2007)~\cite{MAID} solutions are shown in red, gold, blue, and green, respectively.
\label{fig:obrF}}
\end{figure*}

The results of this analysis agree with previous measurements in the overlapping energy range. The major difference between the results of this analysis and the results from previous measurements is in the coverage of the energy range. In this analysis, the upper limit of $W$ is much larger, up to 2525~MeV, compared to $W$ up to 1895 MeV for the observable $F$ from the recent measurements from MAMI~\cite{AN} and compared to $W$ up to 2285 MeV for the observable $T$ from the recent measurements from CBELSA/TAPS~\cite{HA2}. The precision of this analysis is better than the existing measurements in the range of $W$ above 2100 MeV. The SAID and J{\"u}Bo groups have included the present result in the databases. The full results of the polarization observables $T$ and $F$ are shown in Figs.~\ref{fig:obrT} and ~\ref{fig:obrF}.

Exploratory fits have been made with partial-wave analyses. The SAID partial-wave analysis is a data-driven PWA solution and had previously updated the solution to the 2022 version (SM22)~\cite{sm22said}. The prediction of both $T$ and $F$ agree with the data quite well at lower $W$ below 2100 MeV. For the observable $F$, small deviations are observed at higher $W$ above 2100 MeV. For the observable $T$, relatively large deviations are observed starting from $W$ at 2100 MeV. In this energy range, there were previous measurements available from CBELSA/TAPS such as Ref.~\cite{HA2} and the results of this analysis agree well with them in the overlapped energy range up to about 2285~MeV in $W$. A new solution with the present result included (KX23) has been made. This agrees with this data quite well in both lower and higher $W$ ranges. Table~\ref{tab:said} illustrates how the present data differs from either solution. Since they were data-driven, large $\chi^2$ would be possible. The fit with the present data reduced that significantly.
\begin{table}[hbt!]
\caption{\label{tab:said}%
Deviations of the result compared with solutions.}
  \begin{ruledtabular}
\begin{tabular}{ccccccccc}
$W$ & Avg. $\chi^2/\text{data}$ (SM22) & Avg. $\chi^2/\text{data}$ (KX23) \\
\hline
$<2.1$ GeV  ($T$) & 3.94 & 3.25 \\
$>2.1$ GeV  ($T$) & 10.06 & 2.79 \\
$<2.1$ GeV  ($F$) & 1.97 & 1.48 \\
$>2.1$ GeV  ($F$) & 5.72 & 2.46 \\
\end{tabular}
\end{ruledtabular}
\end{table}

The new data on $T$ and $F$ were also included in the J\"ulich-Bonn analysis and a full re-fit of the free model parameters was performed that improved the description of the data at higher energies compared to the predictions of the J{\"u}Bo (2022) solution~\cite{Jubo}, especially for $W$ above 1900 MeV. The J{\"u}Bo model~\cite{ronchen2013coupled}~\cite{ronchen2014photocouplings} is a dynamical coupled-channel approach that preserves unitarity and analyticity and aims at the extraction of the light baryon resonance spectrum from a simultaneous analysis of the reactions $\pi N$, $\gamma p\to \pi N$, $\eta N$, $K\Lambda$ and $K\Sigma$. A Lippmann-Schwinger equation formulated in time-ordered perturbation theory (TOPT) is solved including off-shell intermediate momenta and the resonance poles $W_0$ are determined on the unphysical Riemann-sheet of the $T$-matrix.

The new data have a noticeable impact on the extracted resonance parameters. While the pole positions of most of the lower lying well-established states are relatively stable, we find significant changes in the mass and width of the $\Delta(1910)1/2^+$. The new mass is 1748 MeV compared to 1802 MeV in  J{\"u}Bo (2022). The state is also much more narrow in the re-fit: 353 MeV compared to a width of 550 MeV in J{\"u}Bo (2022). It should be noted that this state has been rather unstable in various  J{\"u}Bo fits and also the range of the mass and width listed by the PDG is comparatively large ($Re [W_0]=1860\pm30$~MeV, $-2Im [W_0]=300\pm100$~MeV). The situation is similar for the $N(2190)7/2^-$ for which a significant change in the imaginary part of the pole position is observed in the re-fit: $W_0=1950-i78$~MeV compared to $W_0=1965-i144$~MeV in J{\"u}Bo (2022). Also this state showed varying values of the resonance parameters in different J{\"u}Bo analyses. New data, especially on polarization observables as the present ones, are very helpful to fix the parameters of the $\Delta(1910)1/2^+$ and $N(2190)7/2^-$.

Besides the changes in the pole positions of the latter two states, we find significant differences in the photocouplings at the pole for the $\Delta(1930)5/2^-$, $\Delta(2200)7/2^-$ and $N(2220)9/2^+$. For broad states in higher partial waves those quantities are difficult to determine in general. That applies especially to the isospin $I=3/2$ $\Delta$ states. New data from mixed-isospin channels as $\gamma p\to\pi^0p$ therefore provide valuable input to fix those quantities.

\section{Summary}

Polarization observables $T$ and $F$ in the $p(\gamma,\pi^0)p$ reaction have been extracted for the center-of-mass energy from 1.445 GeV to 2.525 GeV in the FROST experiment with the CLAS detector at JLab. The results of observables $T$ and $F$ in this analysis agree with previous measurements in the overlapping energy range but have a much larger coverage at higher energies. The present SAID, BnGa, MAID, and J{\"u}Bo model predictions generally agree with the data in the energy range where previous measurements exist but significant deviations are observed at higher energies. The polarization observables, especially in the energy range that was not included in previous measurements, becomes useful for the partial-wave analysis to fulfill the completeness in the determination of the pion-photoproduction reaction and to study this reaction channel in a much larger energy range. The present data has been incorporated into the databases of the partial-wave analysis groups. The SAID fits have been improved with relatively small $\chi^2$. Significant changes in the parameters of the $\Delta(1910)1/2^+$ and $N(2190)7/2^-$ have been found with the J{\"u}Bo model.

\begin{acknowledgments}
The authors gratefully acknowledge the efforts of the staff in the Accelerator and the Physics Divisions at Jefferson Lab. This work was supported in parts by the U.S. National Science Foundation (NSF) Award No. PHY-1505615, U.K. Science and Technology Facilities Council (STFC) Grant No. ST/V00106X/1, U.S. Department of Energy, Office of Science, Office of Nuclear Physics, under Awards No. DE–SC0016583 and DE–SC0016582, the Deutsche Forschungsgemeinschaft (DFG) from Project- ID 196253076-TRR 110. The authors also gratefully acknowledge the computing time on the supercomputer JURECA at Forschungszentrum J{\"u}lich under Grant No. “baryonspectro”. This work was also supported by the Italian Istituto Nazionale di Fisica Nucleare (INFN), the French Centre National de la Recherche Scientifique (CNRS), the French Commissariat {\`a} l’{\'E}nergie Atomique (CEA), the Deutsche Forschungsgemeinschaft (DFG), the National Research Foundation of Korea (NRF). Jefferson Science Associates, LLC, operates the Thomas Jefferson National Accelerator Facility for the U.S. Department of Energy under Contract No. DE-AC05-06OR23177.
\end{acknowledgments}

\bibliography{pi0p}

\end{document}